\documentclass{aa}  
\usepackage{graphicx}
\usepackage{txfonts}
\begin{document}
   \title{On the age of the TW Hydrae Association and 2M1207334$-$393254}

%   \subtitle{Accurate chronology }

   \author{David Barrado y Navascu\'es
          \inst{1}\thanks{Based on observations collected with the Magellan Telescopes}
          }

   \offprints{D. Barrado y Navascu\'es}

   \institute{Laboratorio de Astrof\'{\i}sica Espacial y F\'{\i}sica Fundamental,
LAEFF-INTA, P.O. Box 50727, E-28080 Madrid, SPAIN\\
              \email{barrado@laeff.inta.es}
             }

   \date{Received September 15, 1996; accepted March 16, 1997}

% \abstract{}{}{}{}{} 
% 5 {} token are mandatory
 
  \abstract
   {}
   {We have estimated the age of the young moving group TW Hydrae Association, a cohort 
    of a few dozen stars and brown dwarfs located near the Sun which 
    share the same kinematic properties and,
    presumably, the same origin and age.}
   {The chronology has been determined
    by analyzing different properties (magnitudes, colors, activity, lithium) 
    of its members and comparing them with several well-known
    star forming regions and open clusters, as well as theoretical models.
    In addition, by using medium-resolution optical spectra
    of two M8 members of the association (2M1139 and 2M1207 --an accreting brown dwarf with a 
    planetary mass companion), we have 
    derived spectral types and measured  H$\alpha$ and lithium equivalent widths. We have also 
    estimated their     effective temperature and gravity, which were used to produce 
    an independent age estimation for these two brown dwarfs.
    We have also  collected spectra of  2M1315, a candidate member with a L5 spectral type and
    measured its H$\alpha$ equivalent width.}
   {Our age estimate for the association, 10$^{+10}_{-7}$ Myr, agrees with previous values
    cited in the literature. In the case of the two brown dwarfs, we have derived
    an age of  15$^{+15}_{-10}$ Myr,  which also agree with our estimate for the whole  group.}
   {We compared our results with recent articles published on the same subject using other techniques, 
    and  discuss the limits of the age-dating techniques.}


   \titlerunning{The age of TWA and 2M1207}
   \authorrunning{David Barrado y Navascu\'es}

   \maketitle
%
%________________________________________________________________

\section{Introduction}

%%%%%%%%%%%%%%%%%%%%%%%%%%%%%%%%%%%%%
%%%%%%%%%%%%%%%%%%%%%%%%%%%%%%%%%%%%%
%%%%%%%%%%%%%%%%%%%%%%%%%%%%%%%%%%%%%
%%%%%%%%%%%%%%%%%%%%%%%%%%%%%%%%%%%%%
%%%%%%%%%%%%%%%%%%%%%%%%%%%%%%%%%%%%%
%%%%%%%%%%%%%%%%%%%%%%%%%%%%%%%%%%%%%
%%%%%%%%%%%%%%%%%%%%%%%%%%%%%%%%%%%%%

If we want to put any astronomical phenomenon in context, 
we need to know its evolutionary status, i.e. its age. This is
even more important when the phenomenon is a benchmark in a given discipline. This is the
case  of the TW Hydrae Association or TWA (de la Reza et al.1989; Kastner et al. 1997),
 a young, nearby moving group
which includes a few dozen stars and brown dwarfs 
(see the recent review by Zuckerman \& Song 2004).
The age has been estimated  in the range 8-20 Myr (Kastner et al. 1997; 
Stauffer,  Hartmann \& Barrado y Navascu\'es 1995; Soderblom et al.  1998;
Hoff et al. 1998; Weintraub  et al. 2000; Makarov \& Fabricius 2001; 
Makarov et al. 2005;
de la Reza, Jilinski, Ortega 2006),
 and the individual  distances,  from Hipparcos, are  in the range 47-67 pc (ESA 1987).
Due to these  characteristics,  the TWA is becoming a reference for such phenomena
as accretion and circumstellar disks, rotation and activity,  the formation
of planetary systems and the direct detection of planets. 

Regarding the latter issue,  members of the association have been the targets 
of several searches for planetary companions, using AO techniques. Of those studies 
which have been  published,  
the most comprehensive one has been that by Masciadri et al. (2004), who observed 
30 targets with VLT/NACO, including six stellar members  of TWA, with no positive result.
However,  Chauvin et al.  (2004) announced the discovery of a candidate
companion of the brown dwarf 2M1207334-393254 (2M1207 hereafter), which has a mass estimate
within the planetary domain (5 M$_{jup}$).  One year later, the same team (Chauvin et al. 2005) 
measured the proper  motion and  verified that the brown dwarf and the companion were moving
together, confirming  the nature of this object, the first planet whose detection
 had been achieved by direct imaging.
 
The 2M1207 system is full of surprises, since the brown dwarf is still
undergoing active accretion as shown by Mohanty, Jayawardhana \& Barrado y Navascu\'es (2004)
from high--resolution optical spectra around the H$\alpha$ lines. It harbors a  circumsubstellar disk, 
seen as a mid-IR excess by Sterzik et al. (2004), which can 
be explained   by optically thick dust disk models. 
Rioz, Gizis \& Hmiel (2006) have also derived the IR excess by using Spitzer data.
However, no radio emission has been detected (Osten \& Jayawardhana 2006).
Mohanty (2006, priv. comm.) has suggested that the secondary --the planetary mass companion-- 
might have a disk too.
Age is a key parameter in  understanding the nature of the faint companion and 
the properties of the system.

In the present analysis, we will present in section 2 the new spectroscopic data
collected for several among the lowest mass candidate 
members of the association (two M8 and a L5), together
with an estimate of the age and masses. 
Section 3 will  revisit the data regarding the
known members of the association, to estimate an accurate age by using different 
methods. We will end the paper with the conclusions and a discussion about the limits of
 the age-estimate techniques.

\section{Two M8 and a L5: 2M1139, 2M1207 and 2M1315}
%%%%%%%%%%%%%%%%%%%%%%%%%%%%%%%%%%%%%
%%%%%%%%%%%%%%%%%%%%%%%%%%%%%%%%%%%%%
%%%%%%%%%%%%%%%%%%%%%%%%%%%%%%%%%%%%%
%%%%%%%%%%%%%%%%%%%%%%%%%%%%%%%%%%%%%
%%%%%%%%%%%%%%%%%%%%%%%%%%%%%%%%%%%%%
%%%%%%%%%%%%%%%%%%%%%%%%%%%%%%%%%%%%%
%%%%%%%%%%%%%%%%%%%%%%%%%%%%%%%%%%%%%

\subsection{New spectroscopic observations: spectral classification}
%%%%%%%%%%%%%%%%%%%%%%%%%%%%%%%%%%%%%
%%%%%%%%%%%%%%%%%%%%%%%%%%%%%%%%%%%%%

We collected medium-resolution spectra of three low mass candidate members
of TWA, namely 
2M1139511$-$315921 (hereafter 2M1139), 
2M1207433$-$393254 (2M1207), and 
2M1315309$-$264951 (2M1315), 
with the Magellan/Baade telescope and the Boller \& Chivens (B\&C)
 spectrograph on March 11, 2003. We used the 1200 l/m gratings.
The achieved spectral resolution is R=2600, as measured in the 
comparison arcs, with a  spectral coverage of 6200--7800 \AA.
During the observing run,
 among other targets, we observed 2M1139, 2M1207 and 2M1315, candidate members of the 
TW Hydrae association discovered by Gizis (2002) --see also Hall (2002a) in the case of 2M1315.
 The first two have been classified
as M8 spectral type, whereas 2M1315 is much cooler, with a spectrum corresponding to L5. 
We collected six consecutive spectra of 2M1139, 600 seconds each;
four of 2M1207, with the same exposure time, and seven of 2M1315,
the first one of 300  seconds and the others of 1200 seconds each.
The  data were reduced using standard techniques within 
IRAF\footnote{IRAF is distributed by National Optical Astronomy Observatories, 
which is operated by the Association of Universities for Research in Astronomy, Inc.,
 under contract to the National Science Foundation, USA}.  
The individual spectra were combined using median algorithm in order to remove the 
cosmic rays.
The final spectra can be seen in Figure 1.

In addition to these objects, we observed a large number of spectral templates
with spectral types in the interval K7-M9, both giants and dwarfs.
More details regarding the observations and the reduction process can be obtained
in Barrado y Navascu\'es, Mohanty  \& Jayawardhana (2004), 
Barrado y Navascu\'es \& Jayawardhana (2004) and  
Barrado y Navascu\'es et al. (2004).
For high-resolution spectroscopy of the two M8 targets,
 taken with Magellan/Baade and the echelle spectrograph 
MIKE, see Mohanty, Jayawardhana \& Barrado y Navascu\'es (2003).

Based on the comparison with the spectral templates 
(namely  HIP70669 --M6e--,  vB07 --M7 V--, HIP68137 --M7--,
HIP63642 --M8III--, LHS2397 --M8V--, and BRI-0337-35 --M9),
we  have derived   spectral types for 2M1139 and 2M1207.
As can bee seen in Figure 2 (only in the electronic version), where we display all these spectra 
(the solid lines correspond to the templates, whereas the dotted 
lines represent the TWA candidate members), the most similar templates 
are the M8V and M8III,   the first case being closer to matching the target spectra.
For comparison purposes,  the spectra have been normalized in the area
 around 7500 \AA. Therefore, it would be possible to classify these two
objects as M8IV (they are contracting towards the Main Sequence).
 This classification is 
 in  agreement with Gizis (2002), who concluded that they have low gravity
based on the VO indices measured from low-resolution
spectroscopy, and by  Mohanty, Jayawardhana \& Barrado y Navascu\'es (2003)
--from high-resolution spectroscopy.
There are obvious differences between the M8 and M8 III templates
 and our TWA targets.
 See section 2.2.2 for a further analysis 
concerning the luminosity class and the surface gravity.

We have also compared the spectra of 2M1139 and 2M1207 with each other, in other to verify
whether they  belong to the same spectral type. The visual inspection 
indicates that they are very similar, even when over-plotting both spectra.
 However, when the  difference (one minus the other) 
or the ratio (one divided by the other) of both spectra 
is computed,  2M1207 might be slightly cooler that 2M1139.
Figure 1b displays the difference between both spectra (2M1207$-$2M1139), 
after normalizing them using the flux at 7000 \AA.
As a reference, we have plotted in the bottom panel the spectrum of 2M1139 
and both panels indicate the location of key spectral features, such as the headband 
of TiO or the central wavelength of the VO. Two characteristics are evident from 
this graph: First, the flux of 2M1207 is greater than 2M1139 both in the red and blue
part of the spectral range. This might be due to veiling --as clearly shown by 
Mohanty, Jayawardhana \& Barrado y Navascu\'es (2003), 2M1207 is weakly accreting--
and to a cooler spectral type. Second, the VO band is substantially different in both cases, 
being shallower at the central part and deeper at the edges in the case of 2M1207.
Conversely, this can be interpreted as 2M1207 having a wider, shallower VO band.
This might be related to differences in gravity (objects not exactly coeval) or temperature.

\subsection{H$\alpha$ emission}
%%%%%%%%%%%%%%%%%%%%%%%%%%%%%%%%%%%%%
%%%%%%%%%%%%%%%%%%%%%%%%%%%%%%%%%%%%%

Since we have collected several consecutive spectra for each of these three candidate 
members of the TW Hydrae Association, it is possible in principle to search for
variability in the H$\alpha$ profile or equivalent width. This variability, if present,
 might be due to activity (both via  modulation by the rotation or because of flares)
or to accretion. We have not found any unambiguous evidence of variability in the case of 2M1139 
or 2M1207. However,  the range covered in time is very short, being only 
a fraction of the expected rotation period for this type of object.
The measured equivalent widths are W(H$\alpha$) = 
10.2$\pm$0.7 \AA{ } and 44.7$\pm$2.0 \AA, respectively.
Previous measurements were 10.0 and  300 \AA, respectively,  in low-resolution 
spectra (Gizis 2002), and 7.3 and 27.7 \AA, 
for 2M1139 and 2M1207, in high-resolution 
spectra (Mohanty, Jayawardhana \& Barrado y Navascu\'es 2003). 
More recently, Gizis \& Bharat (2004)  
measured an equivalent width of 42 \AA{ } in the case of 2M1207. 
They speculate  that Gizis (2002)
 collected the 2M1207 spectrum during a flare.
However, although the flare might have taken place,
 these differences  are to be expected, since, as noted by 
White \& Basri et al. (2003), the higher the
 resolution, the smaller the equivalent width
even if there are not intrinsic differences (more or less a factor of three going from 
low- to high- resolution).
Scholz, Jayawardhana \& Brandeker (2005), by collecting a large number
of high resolution spectra of 2M1207 in two different epochs separated by several weeks,
 have shown that  the profile of the H$\alpha$ line can change dramatically. The actual values
can be found in Scholz \& Jayawardhana (2006).

2M1315, the L5 spectral type  object, with a high proper motion and a
 possible age of more than 3 Gyr, is  likely not a member of
 the association, as discussed by Hall (2002). Since it was observed during the
 same campaign as the two M8  TWA members, we have included the data. 

In the case of 2M1315, some H$\alpha$  variability  might be present, although with some caveats.
The continuum in this region,  corresponding to its spectral type  --L5--, of this  brown dwarf
 is very low, and the measurement of the  equivalent width has a large uncertainty.
The inset in Figure 1a  displays the H$\alpha$ profiles for the seven independent spectra we have 
collected. The spectra have been normalized using the flux in the range around 
7000 \AA{} (where the flux is maximum in the covered spectral range). 
The exposure times are 20 minutes except in the first spectrum, which is only 5 minutes 
long. The total time between the exposures is about 2.5 hours.
As shown by Mohanty \& Basri (2003) and Zapatero Osorio et al. (2006), typical
projected  rotational velocities for L5 are in the range 20-40 km/s.
 A L5 object has an effective temperature of about 1750 K (Basri et al. 2000), and
for this temperature COND models from the Lyon group (Chabrier et al. 2000) are more
 appropriate.  Depending on the assumed age (10 Myr or 5 Gyr), the expected rotation 
period is in the range 2.5--10.5 hours (being shorter for older, faster objects).
 Therefore, we have covered just one rotation period in the optimal case,
and we do not see any obvious modulation, although the data are too sparse.

The average equivalent width is W(H$\alpha$)= 153$\pm$26 \AA. Gizis (2002) measured 
W(H$\alpha$)=97 \AA{ } in a low-resolution spectrum, and Hall (2002ab) has reported 
values of 25, 121 and 124 \AA. The data suggest that we are dealing with a
low mass object with frequent flaring, although other scenarios, such as acretion 
in a much younger object,  cannot be ruled out.

\subsection{The age of 2M1139 and 2M1207}
%%%%%%%%%%%%%%%%%%%%%%%%%%%%%%%%%%%%%
%%%%%%%%%%%%%%%%%%%%%%%%%%%%%%%%%%%%%

\subsubsection{The lithium depletion and the age}
%%%%%%%%%%%%%%%%%%%%%%%%%%%%%%%%%%%%%

Both 2M1139 and 2M1207 show the lithium doublet at 6708 \AA, as can
be seen in Figure 3. In addition, HeI6678 \AA, another
accretion indicator, can be clearly seen in the second object,
with an equivalent width of W(HeI)=0.92$\pm$0.15 \AA.

The measured lithium equivalent widths are W(LiI) = 0.62$\pm$0.13 and
0.51$\pm$0.09 \AA, for 2M1139 and 2M1207, respectively.
  The comparison of these values
with those of similar objects belonging to other young clusters or 
star forming regions (SFRs) indicate that  this element has not been depleted so far
in the two TW Hydra candidate  members. In  principle, depending on the
effective temperature scale and the theoretical models, this fact imposes
a maximum age of about $\sim$50 Myr. 
The lithium abundance can be derived  using the curve of growth
(for example Zapatero Osorio et al. 2002),
the measured equivalent width   and 
an estimate of the  effective  temperature
(i.e., Luhman et al. 1999, 2720 K for M8 with luminosity class in between
III and IV). The derived abundances are
 A(Li)=3.10 and 2.18 for 2M1139 and 2M1207, respectively
--on a scale where  A(Li)=12$+$Log[N(Li)/N(H)].
 The low abundance of 2M1207 can be interpreted as the
presence of moderate veiling, due to accretion. 
Further details  on the  lithium depletion, the spread of the estimated abundance
 and age  are given in section 3.5.

Our spectrum 
of the L5 2M1315 is not good enough  to even detect unambiguously the lithium 
feature, although it must be present due to its spectral type and 
estimated mass range, even if it does not belong to the association and it is much older.
 According to models by Chabrier et al. (2000), it should 
have a mass in the range 0.01-0.07 M$_\odot$, for ages ranging from 10 Myr to 5 Gyr.
In any case, lithium cannot be used to establish the membership status of 2M1315
or to derive an age estimate for it.

\subsubsection{The surface gravity, the mass and the age}
%%%%%%%%%%%%%%%%%%%%%%%%%%%%%%%%%%%%%

The distances to 2M1139 and 2M1207, to the best of our knowledge, 
 have not been measured so far by obtaining trigonometric parallaxes. 
Therefore, it is not possible to 
locate these objects in color-color or HR diagrams, without any
assumption regarding their distance.
However, Mamajek (2005) has derived a distance based on the moving cluster method.

It is possible to use the effective temperature 
and the surface gravity to estimate both the age and 
the mass of these two objects. This method is independent of the distance and
does not require one  to prove that they belong to the TW Hydrae Association. 
Figure 4 (only in the electronic version) displays a comparison between the spectra of 2M1139 and 2M1207 with several
low mass stars in the spectral range around 7500 \AA. These range  includes 
a VO and a TiO bands and the KI7700 doublet.  This alkali feature is gravity sensitive 
and can be used to estimate the surface gravity. Based on the diagram,  
the closest templates are HIP63642 (M8III), LHS2397a (M8V) and BRI-0337-35 (M9).
However, small differences persist for both the continuum and the KI7700 doublet.
We have estimated that the luminosity class  has to be in between III and V, 
with an intermediate  gravity, somewhat closer to V than to III 
(logg=4.25$\pm$0.5).

The comparison of the derived surface gravity and effective temperature
with theoretical models is presented in  Figure 5 
(solid and dashed lines  for isochrones and tracks, respectively).
The models correspond to Baraffe et al. (1998) and the effective temperature
to the average between giants and dwarfs by Luhman (1999). Note that
other scales, such as dwarfs by Bessell (1991),  Leggett (1992), or Luhman (1999), 
would give  temperatures for 2M1139 and 2M1207 about 100-150 K cooler, providing
smaller masses and ages slightly older. However, the temperature scale by
 Luhman (1999) has been derived to match the Baraffe et al. (1998) models.
Figure 5 indicates that the masses of these two objects are 0.04$\pm$0.01 M$_\odot$, and
the age slightly younger than 15 Myr, although the uncertainty in the derived surface 
gravity allows for much younger or older ages (from 1 to 30 Myr); a  
more realistic lower limit would be 5 Myr. 

Spectral synthesis in young brown dwarfs
 produces effective temperatures  about 200 K cooler
 than the values derived from the scale of Luhman (1999). An example corresponding to the
young brown dwarf LS-RCrA 1 can be found in
Barrado y Navascu\'es, Mohanty \& Jayawardhana (2004).
 In the case of 2M1207, a
effective temperature of about 2550 K can be derived (Mohanty, priv.comm). 
The same analysis produces a value of the gravity of logg=4.0$\pm$0.5. These quantities
would locate 2M1207 on top of younger isochrones, with a maximum age of 25 Myr and a 
probable age of 8 Myr.

The figure shows that these two objects
 are bona fide young brown dwarfs,
independently of their belonging  to the TW Hydrae Association.

\section{Revisiting the TW Hya Association.}
%%%%%%%%%%%%%%%%%%%%%%%%%%%%%%%%%%%%%
%%%%%%%%%%%%%%%%%%%%%%%%%%%%%%%%%%%%%
%%%%%%%%%%%%%%%%%%%%%%%%%%%%%%%%%%%%%
%%%%%%%%%%%%%%%%%%%%%%%%%%%%%%%%%%%%%
%%%%%%%%%%%%%%%%%%%%%%%%%%%%%%%%%%%%%
%%%%%%%%%%%%%%%%%%%%%%%%%%%%%%%%%%%%%

\subsection{The data}
%%%%%%%%%%%%%%%%%%%%%%%%%%%%%%%%%%%%%
%%%%%%%%%%%%%%%%%%%%%%%%%%%%%%%%%%%%%

One of the goals of this paper is to derive an
 accurate age for the TW Hydrae Association.
To do so, we have collected data from the literature.
So far, there are a few dozen candidate members of the association.
 However, we have restricted ourselves to those whose parallaxes
have been measured, or whose properties have been studied 
in depth (lithium and  H$\alpha$ equivalent widths, and so on).
There are several age indicators that  have been used, such as
the rotational velocity, the orbital period or the coronal activity
 as measured from  X-rays. However, they have a very large scatter 
for a particular spectral type and they cannot provide an age estimate accurate  enough 
for our purposes. For this reason, we have only studied  the 
location in a HR diagram, the lithium and the H$\alpha$ emission.

Among the papers we have used to extract data are:
Rucinski \& Krautter   (1983),
de la Reza et al.      (1989),       
Jura et al.            (1993),        
Stauffer et al.        (1995),
Soderblom et al.       (1996),
Hoff et al.            (1998),
Sterzik  et al.        (1999),      
Webb et al.            (1999),               
Neuhauser et al.       (2000),      
Torres et al.          (2000),      
Muzerolle, priv. comm  (2000),   
Zuckerman et al.       (2001),       
Gizis                  (2002),          
Song et al.            (2002),        
Mohanty et al.         (2003),       
Reid et al.            (2003),           
Song et al.            (2003),        
Torres et al.          (2003),        
Gizis et al.           (2004),       
Weinberger et al.      (2004),      
Zuckerman \&  Song     (2004),      
Scholz \& Jayawardhana (2006),
Jayawardhana et al.    (2006), and
Scholz et al.          (2006).
In this paper we will use Hipparcos distances (ESA 1987).
 However,  Mamajek (2005), using the moving cluster method, has been able to derive
an individual distance for 2M1207 (56$\pm$6 pc)	and for the association as a whole
(49 pc). The data are listed in Table 1.

The bolometric luminosities were derived from the magnitudes in different
bands (namely $V$, $I$, $J$ and $Ks$), the Hipparcos parallax and the
bolometric corrections for each band.
The latter values  were computed using 
the expressions provided by 
 Schmidt-Kaler and the spectral type for $V$;
Bessell (1991) and the color index ($I-K$) for $I$;
Lawson (1996) and the spectral type for $J$; and
Tinney (1993) and the color index ($I-K$) for $Ks$.

M$_{bol}$ = 5 + 5 Log $\pi$ + M(band) + BC(band)

L$_{bol}$ = 10$^{-0.4 (M_{bol} - M_{bol}^\odot)}$ (L/L$_\odot$)

where M$_{bol}$$^\odot$  is the bolometric magnitude of the Sun and 
has been assumed to be 4.75 mag. Errors in the bolometric luminosity
were estimated adding in quadrature those produced by the errors
in the distance and those coming from either the photometry 
--for the individual values coming from
each band-- or the dispersion from the average of those  values.

\subsection{The Color-Magnitude diagram: membership 
and an initial age estimate}
%%%%%%%%%%%%%%%%%%%%%%%%%%%%%%%%%%%%%
%%%%%%%%%%%%%%%%%%%%%%%%%%%%%%%%%%%%%

Figure 6 represents an infrared  Color-Magnitude Diagram (CMD),
 and includes evolutionary tracks and isochrones by Baraffe et al. (1998),
as well as the location of several TW Hydrae candidate members whose
parallaxes have been measured.
The $J$ and $Ks$ magnitudes come from the 2MASS All Sky Survey (Cutri et al. 2003).
Although IR CMDs are not the best ones to derive masses and ages, 
the optical photometry of these objects was not derived in a homogeneous way or
 simultaneously for all members. Therefore, we have preferred to use the 2MASS data
as a first approach to asses the membership of this handful of candidates.
Errorbars include the error coming from the photometry and the distance, added
in quadrature.
This diagram clearly indicates that HIP53486 and HIP50796, which have been 
proposed by Makarov \& Fabricius (2001), Song et al. (2002) and
Torres et al. (2003), are not members of the association, being too old.
Moreover, the visual binary TWA09 might not belong to the association either
(see next subsection). Based on this figure, an age about 3-8 Myr can be estimated.
As noted by the authors, colors for the Baraffe et al. (1998) and other 
theoretical isochrones are not certain. They recommend the use of 
bolometric luminosities instead.

\subsection{The Hertzprung-Russell Diagram: an accurate age}
%%%%%%%%%%%%%%%%%%%%%%%%%%%%%%%%%%%%%
%%%%%%%%%%%%%%%%%%%%%%%%%%%%%%%%%%%%%

The location of those TW Hydrae members in a HR Diagram is presented
in Figures 7, 8 and 9 (only in the electronic version). In the case of the four  panels of Figure 7, we have  
displayed models by Baraffe et al. (1998);
the  effective temperature was derived using the
scale by Luhman (1999) (average between III and V luminosity classes), but in each
panel the bolometric luminosity was derived using a different method
(see section 3.1). Thus, they provide an estimate of the error
introduced by the conversion from magnitudes into luminosities.
In the case of the three panels of Figure 8, we have only changed the
temperature scale, and  we have included values derived from scales by
Bessell (1991), Leggett (1992) and  Luhman (1999).
Figure 9 provides the effect on the age of different  theoretical models.
Regardless of the specific details (the model, the temperature or the method used 
to derive luminosities), we believe that the age is well constrained by the 
3 and 20 Myr isochrones, with an optimal  value of 8 Myr.

The location of TWA09 seems to indicate that this object is a spurious
member of the association. This result agrees with recent estimates by de la Reza, Jilinski 
\& Ortega (2006), who studied the dynamical evolution of the association, showing 
that it has a different  orbit around the Galaxy
compared to  other members.

\subsection{The H$\alpha$ emission, the fraction of accreting members and the age.}
%%%%%%%%%%%%%%%%%%%%%%%%%%%%%%%%%%%%%
%%%%%%%%%%%%%%%%%%%%%%%%%%%%%%%%%%%%%

In Barrado y Navascu\'es \& Mart\'{\i}n (2003) --see also
 Barrado y Navascu\'es et al.  (2004)-
we have shown that the H$\alpha$ equivalent width, as measured
using low-resolution spectroscopy, is a powerful criterion from 
the statistical point of view to classify accreting and 
non-accreting objects, since the saturation of the chromospheric activity
 at  log \{Lum(H$\alpha$)/Lum(bol)\}=$-$3.3, which appears in members
of  young open clusters,     provides a clear 
discrimination between Classical TTauri stars (and substellar analogs) 
and Weak-line TTauri stars (or Post TTauri stars).
We have used this criterion with the TW Hydrae Association.
Figure 10 displays the H$\alpha$ equivalent width versus the spectral type.
Solid circles represent low mass stars and brown dwarfs which are
 undergoing accretion and/or having circunstellar disks
 (as measured using high-resolution  spectroscopy and/or
infrared excess), whereas open circles correspond to non-accreting  objects.
Other TWA candidate members, whose membership is dubious, appear
as crosses.  As can be seen, the criterion is fulfilled in all cases: those 
objects having W(H$\alpha$) larger than the saturation limit do have accretion and
those below it do not show any evidence of it.

We have derived the ratio between accreting and the total number 
of members. To avoid statistical problems with  low numbers, we have restricted 
ourselves to the stellar domain. Depending on how the visual and spectroscopic
binaries and multiple systems 
are accounted for and including errors (computed as Poisson noise),
 this ratio is between 0.20 and 0.0625, with an optimum value of 0.13.
We have derived this ratio for several well known star forming regions and very
young clusters (see details in  Barrado y Navascu\'es \& Mart\'{\i}n 2003) and compared 
them with the result for the 
TW Hydrae Association. Figure  11 summarizes this comparison.
The solid, thick line represents the behavior of the SFRs, whereas the best value
and upper/lower limits for the TWA appear as a dashed line and two dotted lines, 
respectively. If we assume that TWA follows the trend seen in the SFRs, we can estimate
the age of the former group as 7$^{+8}_{-3}$ Myr.

\setcounter{table}{1}
\begin{table}
\caption[]{Summary of the age estimates (in Myr).}
\begin{tabular}{lrrr}
\hline
 Method           & Minimum& Age    &  Maximum \\
\hline 
 Log grav$^1$     &      3 &     15 &  30$^2$   \\
 CMD, parallax$^3$&      3 &      5 &   8   \\
 HRD, parallax$^3$&      3 &      8 &  20   \\
 Disk fraction    &      4 &      7 &  15   \\
 W(H$\alpha$)     &      4 &      7 &  15   \\
 W(Li)            &      5 &     10 &  15   \\
 A(Li)            &$\sim$3 &$\sim$8 &  15   \\
\hline
\end{tabular}
\\
$^1$ Only for 2M1139 and 2M1207.\\
$^2$ Based on spectral synthesis, the upper limit for the age is 15 Myr,
with Teff=2550 K\\
$^3$ Only for those having parallax.
\end{table}
%%%%%%%%%%%%%%%%%%%%%
%%%%%%%%%%%%%%%%%%%%%
%%%%%%%%%%%%%%%%%%%%%
%%%%%%%%%%%%%%%%%%%%%
%%%%%%%%%%%%%%%%%%%%%
%%%%%%%%%%%%%%%%%%%%%

\subsection{Lithium depletion and  age}
%%%%%%%%%%%%%%%%%%%%%%%%%%%%%%%%%%%%%
%%%%%%%%%%%%%%%%%%%%%%%%%%%%%%%%%%%%%

Our final criteria to estimate the age of the TWA are the lithium equivalent
width and abundance. First, we have compared the W(LiI) with several young
 clusters and SFRs. For simplicity, we only show in Figure 12 a comparison
with the Lambda Orionis open cluster (Collinder 69). The data come from
Dolan \& Mathieu (1999, 2001) and Barrado y Navascu\'es et al. 
(2006, in preparation, see this paper for details).
In this diagram, the solid line corresponds to the upper envelope of the values
measured in young open clusters (with ages in the range 30-150 Myr).
The long-dashed line delimits the areas for weak-line and post-TTauri
stars (adapted from Mart\'{\i}n 1997 and Mart\'{\i}n \& Magazz\`u 1999).
We have computed the lithium depletion isochrones using the temperatures and 
abundances listed in the models by Baraffe et al. (1998), the curves of growth 
of Zapatero Osorio et al. (2002) and the effective temperature for 
luminosity class IV by Luhman (1999). These lithium isochrones are represented 
as thick, short-dashed lines in the figure (1, 8, 10, 15 and 20 Myr). 
Note that on the 1 Myr isochrone, lithium has not been depleted, and still has 
an abundance of A(Li)=3.1, the cosmic value.
In Figure 12,  TWA members are displayed as solid circles (CTT),
 open circles (WTT/PTT) and crosses (probable non-members).
The asterisks correspond to members of the  Lambda Orionis cluster 
located in the central area  (after Dolan\& Mathieu 1999, 2001;
 and Barrado y Navascu\'es et al. 2004). As shown by
Dolan \& Mathieu (1999, 2001, 2002), the Lambda Orionis cluster
is about 5 Myr. It cannot be older than 8 Myr, since the star $\lambda$$^1$ itself, 
a O8 III, with a mass estimate of about 30 M$_\odot$ 
(Barrado y Navascu\'es, Stauffer \& Bouvier 2004)
 would have exploded as a supernova in that case.

The W(Li) presents a clear scatter for mid-M stars and brown dwarfs 
(this also happens
for other spectral ranges). The origin of this phenomenon is not clear but 
it can be related to rotation, activity, or a low S/N --ie., large errors.
Most of the objects in the Lambda Orionis cluster 
are clustered around the loci of A(Li)=3.1 (i.e., the cosmic abundance, undepleted).
A substantial fraction of TWA members have W(Li)s smaller than the values 
corresponding to the cosmic abundance and might have undergone some depletion.
On the  other hand, some veiling might be present in 2M1207 (see section 2.2.1).
Therefore, this figure indicates that  TWA is older than 
 the Lambda Orionis cluster.
The same result can be achieved with other clusters, such as the Sigma Orionis
cluster (3--5 Myr). The comparison with the theoretical 
lithium depletion isochrones indicate that the TWA is younger that 15 Myr, 
with an average age of 10 Myr.

We have also derived lithium abundances for the TWA members, and  compared 
these values with the prediction from  different models 
(D'Antona \& Mazzitelli 1997; Baraffe et al. 1998; Siess et al. 2000).
To do so we have used the curves of growth of Zapatero Osorio et al. (2002) and
different temperature scales (Bessell 1991; Leggett 1992; Luhman 1999). These
comparisons are presented in nine panels (Figure 13, only in the electronic version). 
Accreting objects, that might
have a W(Li) and therefore a A(Li) affected by  accretion (because of the veiling)
are represented as solid circles, whereas non-accreting objects appear as open circles.
Dubious members are included as crosses. Note the large scatter for any particular
comparison; even the non-accreting objects are far from being in a lithium depletion isochrone.
With some caveats, the age can be estimated as 8 Myr, with lower and upper limits of 3 and 15 Myr.

\section{Discussion and conclusions}
%%%%%%%%%%%%%%%%%%%%%%%%%%%%%%%%%%%%%
%%%%%%%%%%%%%%%%%%%%%%%%%%%%%%%%%%%%%
%%%%%%%%%%%%%%%%%%%%%%%%%%%%%%%%%%%%%
%%%%%%%%%%%%%%%%%%%%%%%%%%%%%%%%%%%%%
%%%%%%%%%%%%%%%%%%%%%%%%%%%%%%%%%%%%%
%%%%%%%%%%%%%%%%%%%%%%%%%%%%%%%%%%%%%

The TW Hydrae association is becoming a cornerstone  in our understanding of phenomena such
as stellar  accretion and the formation of planetary systems, due to the fact that
it contains several dozen candidate members 
(only a fraction of them confirmed), is nearby,
and its age is critical for the evolution of several stellar properties. Our main goal
has been to establish an accurate age for the whole association and, in the process, 
for the brown dwarf 2M1207, an accreting member which contains, at least, a planetary 
mass object located at a projected distance of  about $\sim$50 AU.

We have used the available data for the {\it bona-fide}
  members and compared  them with 
the properties of well-known star forming regions
 and young clusters, as  well as
theoretical models. The methods we have used
 include HR and Color-Magnitude 
Diagrams, surface gravity,  activity, and lithium. 
Table 2 summarizes all the age estimates,
including minimum and maximum values. Depending on 
the method (the property we have used to estimate 
the age), the derived values are in the range of 5 Myr
 (Color-Magnitude Diagram with 2MASS data
and theoretical models, not very reliable) to 15 Myr
 (for the gravity). The ''best'' value is 10 Myr.
The minimum age is imposed by Color-Magnitude 
 and HR Diagrams, as well
as by the lithium abundance, whereas  an upper 
limit of 20 Myr can be derived 
from the HR Diagram. Therefore, we can say 
that the age of the association is 
10$^{+10}_{-7}$ Myr.

In the case of 2M1207, the gravity, as measured
 by mid-resolution optical spectroscopy,
and its comparison with theoretical models by 
Baraffe et al. (1998) indicates
that the age is 15$^{+15}_{-10}$ My, in agreement
 with the age of the association. As
a comparison, Mamajek (2005)  derived an age 
for TWA  based on the moving cluster distance method,
an expansion age. After confirming the membership of 2M1207, 
he  obtained a lower limit of 10 Myr.
 This value is in contradiction with an estimation  based
 on the dynamical evolution of the association
(Makarov \& Fabricius 2001; de la Reza, Jilinski, \& Ortega 2006), 
8.3$\pm0.8$ Myr.  Moreover, Makarov et al. (2005) have found that three  members 
(TWA 1, 4 and 11) might be younger, since they  derived an expansion age of
 4.7$\pm$0.6 Myr for them. However, our own estimation is compatible with
 these  determinations.

In this analysis, we have assumed the coevality of the TW Hydrae Association.
 This is an hypothesis which might not be true. Lawson \& Crause (2005) have found a 
bimodal distribution in the rotation periods, 
and their interpretation is that there are two
distinct populations, which are 10 and 17 Myr old. As we have shown, the 
almost identical 2M1207 and 2M1139 show small differences in their spectrum, which might
be due to different gravities,  which would indicate different  ages. Therefore, non-coevality 
is a possibility. 

We have several  caveats regarding these age estimations.
 First, age estimation is  difficult. 
Each individual method  has its own problems.
Some among them depend on the direct comparison with 
theoretical models (such as the HR diagram),
other include not well-understood properties 
(such as the process of accretion and its evolution
  evolution or the lithium abundance). 
There are also problems transferring observational
 into theoretical quantities, such as from
magnitudes and colors into effective temperatures and luminosities 
(Stauffer, Hartmann \& Barrado y Navascu\'es 1995).
 Moreover, colors can be affected by activity
and related effects (see the  case of 
the Pleiades,  Stauffer et al. 2003).
On the other hand, different methods can provide very different ages.
A case study is lithium depletion chronology for 
young clusters such as NGC2547, IC2391, Alpha Per and the 
Pleiades, which produces ages about 50\% larger 
(Stauffer et al. 1998, 1999, Barrado y Navascu\'es et al.
1998, 1999, 2004; Oliveira et al. 2003)  than 
canonical values, as derived from isochrone 
fitting in the lower Main  Sequence or the turn-over method.

Therefore, our age --or any age-- estimation should 
be  understood in the context   of a particular
age scale, in  an evolutionary sequence. 
The important fact is that we can sort different 
star forming regions,  moving groups, associations
  and open clusters by their age, stating, for
instance, that Taurus is younger than Lambda Orionis,
 which is younger than IC2391. We can be sure that TWA
 is in between these two latter clusters.
Since we have analyzed several associations, trying to 
understand different, although related, 
phenomenology, in the same way (see, for instance, the
 case of protoplanetary disks of $\beta$ Pic and HR4796 and 
Fomalhaut (Barrado y Navascu\'es et al. 1999, 2001 and
 references therein), we  are trying to link
a whole array of properties in a coherent, but relative,
 evolutionary context. However, much work 
remains to be done.

An absolute age-scale (or scales), based on
 homogeneous datasets (same methodology to acquire and analyze  the data), is needed,
coupled with theoretical work, in order to reach an agreement between the estimates provided 
by different properties.

%%%%%%%%%%%%%%%%%%%%%%%%%%%%%%%%%%%%%
%%%%%%%%%%%%%%%%%%%%%%%%%%%%%%%%%%%%%
%%%%%%%%%%%%%%%%%%%%%%%%%%%%%%%%%%%%%
%%%%%%%%%%%%%%%%%%%%%%%%%%%%%%%%%%%%%
%%%%%%%%%%%%%%%%%%%%%%%%%%%%%%%%%%%%%
%%%%%%%%%%%%%%%%%%%%%%%%%%%%%%%%%%%%%

\begin{acknowledgements}
Based on data collected by the Magellan telescopes. 
DByN is indebted to the Spanish
``Programa Ram\'on y Cajal'' and ESP2004-01049 and to R. Jayawardhana and S. Mohanty.
The comments of the referee, M.  Sterzik, have been very helpful.
 This publication makes 
use of data products from the Two Micron All Sky Survey.
\end{acknowledgements}

%%%%%%%%%%%%%%%%%%%%%%%%%%%%%%%%%%%%%%%%%%%%%
%%%%%%%%%%%%%%%%%%%%%%%%%%%%%%%%%%%%%%%%%%%%%
%
%  Figures 
%
%%%%%%%%%%%%%%%%%%%%%%%%%%%%%%%%%%%%%%%%%%%%%
%%%%%%%%%%%%%%%%%%%%%%%%%%%%%%%%%%%%%%%%%%%%%
%%%%%%%%%%%%%%%%%%%%%%%%%%%%%%%%%%%%%%%%%%%%%
%%%%%%%%%%%%%%%%%%%%%%%%%%%%%%%%%%%%%%%%%%%%%
%%%%%%%%%%%%%%%%%%%%%%%%%%%%%%%%%%%%%%%%%%%%%

%-----------------------------------------------------------
    \begin{figure*}
    \centering
    \includegraphics[width=7.8cm]{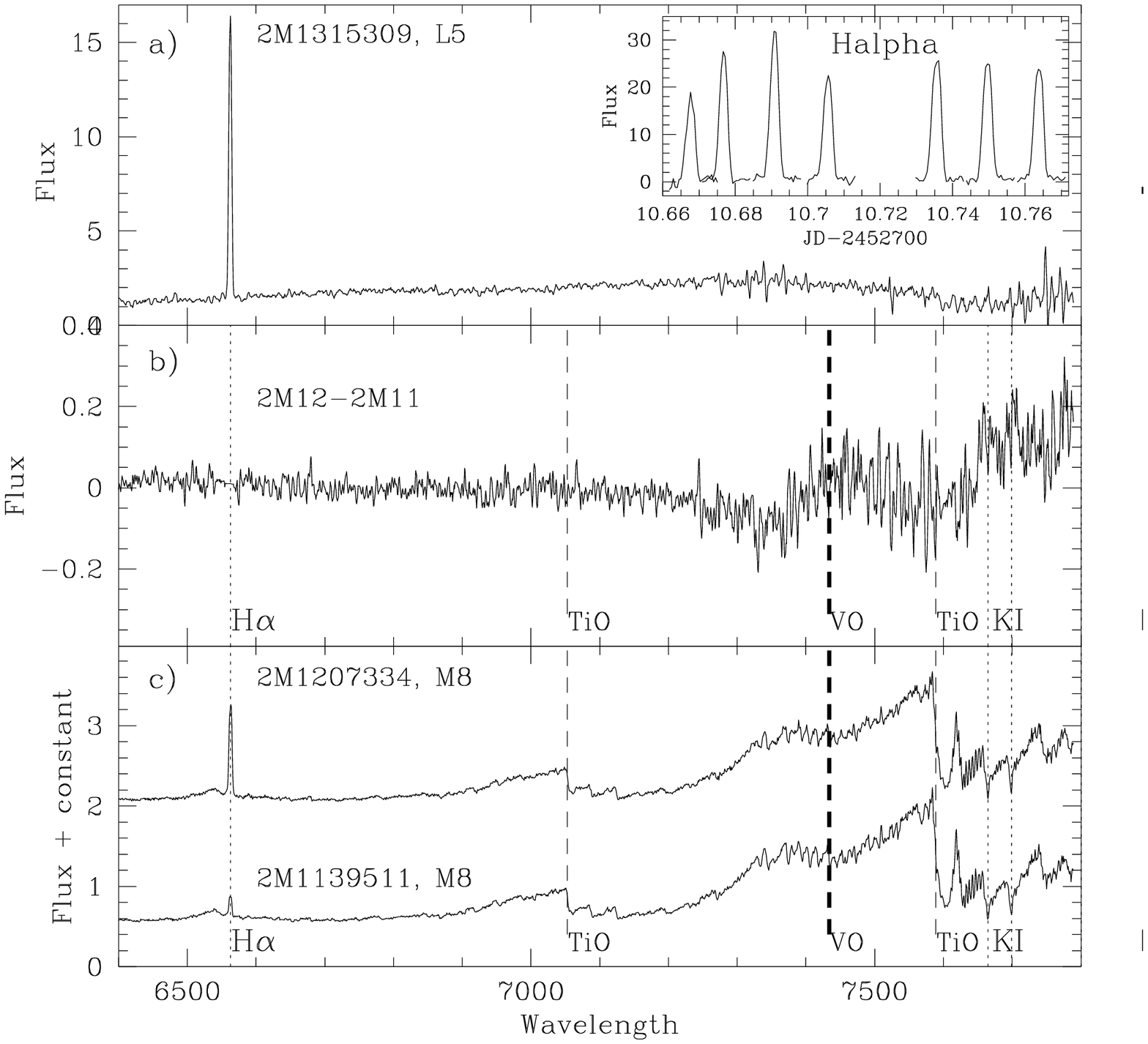}
 \caption{Panel a. Medium-resolution spectrum of 2M1315
Panel c. Medium-resolution spectra of 2M1139 and 2M1207.
Panel b displays the result of the subtraction 2M1207$-$2M1139, after normalization at 7000 \AA.
 Both objects are  very similar, but 2M1207 has a flux excess in the red side 
of the spectrum, suggesting it might be somewhat cooler. Some excess in the 
blue side is apparent, which might be an effect of the accretion present in 2M1207.
The inset in panel a displays the H$\alpha$ profiles of 2M1315 (L5 spectral type).
 The continuum have been normalized using the flux at 7000 \AA.}
 \end{figure*}
%______________________________________________________________      
%%%%%%%%%%%%%%%%%%%%%
%%%%%%%%%%%%%%%%%%%%%

%-----------------------------------------------------------
    \begin{figure*}
    \centering
    \includegraphics[width=7.8cm]{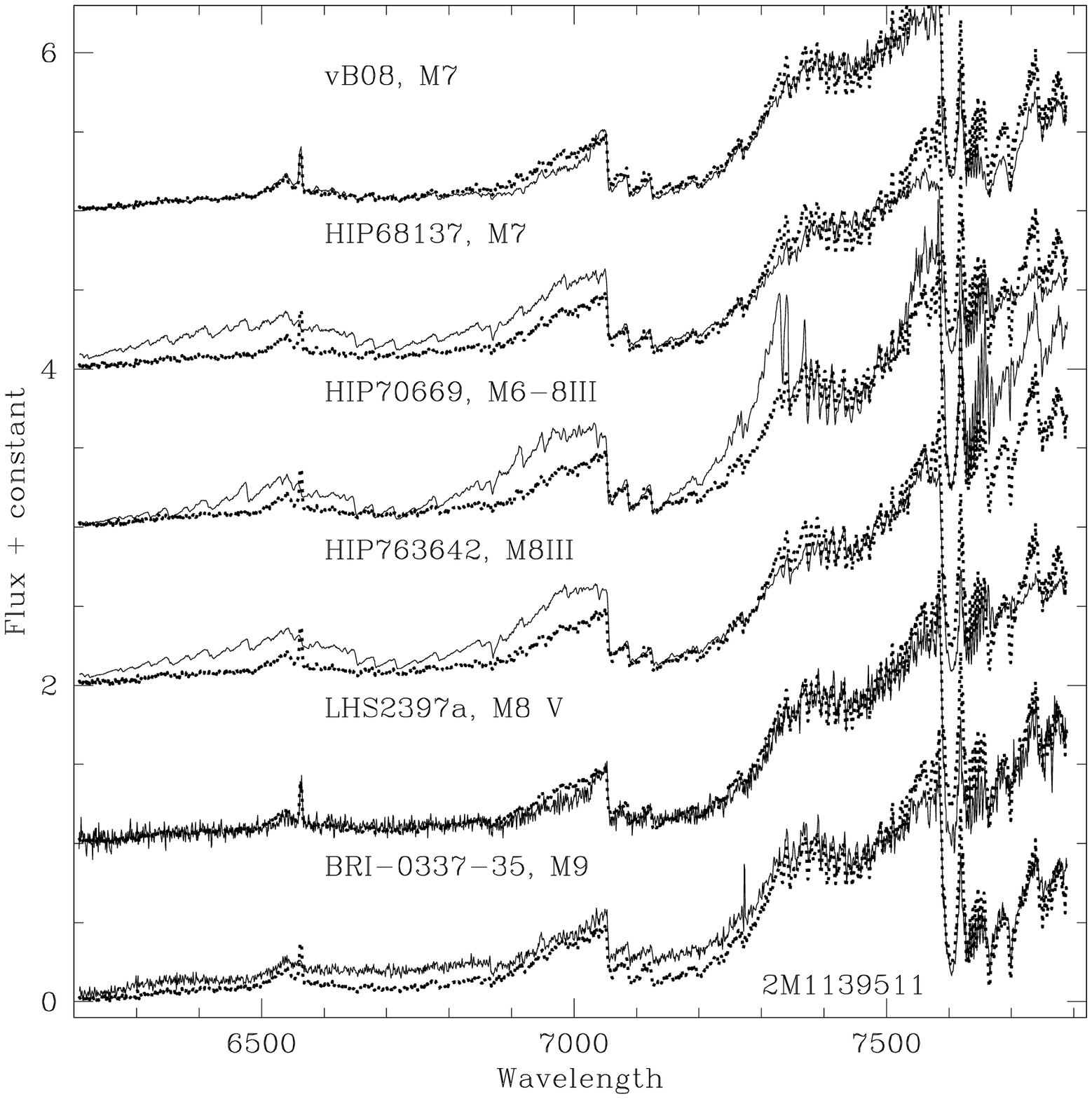}
    \includegraphics[width=7.8cm]{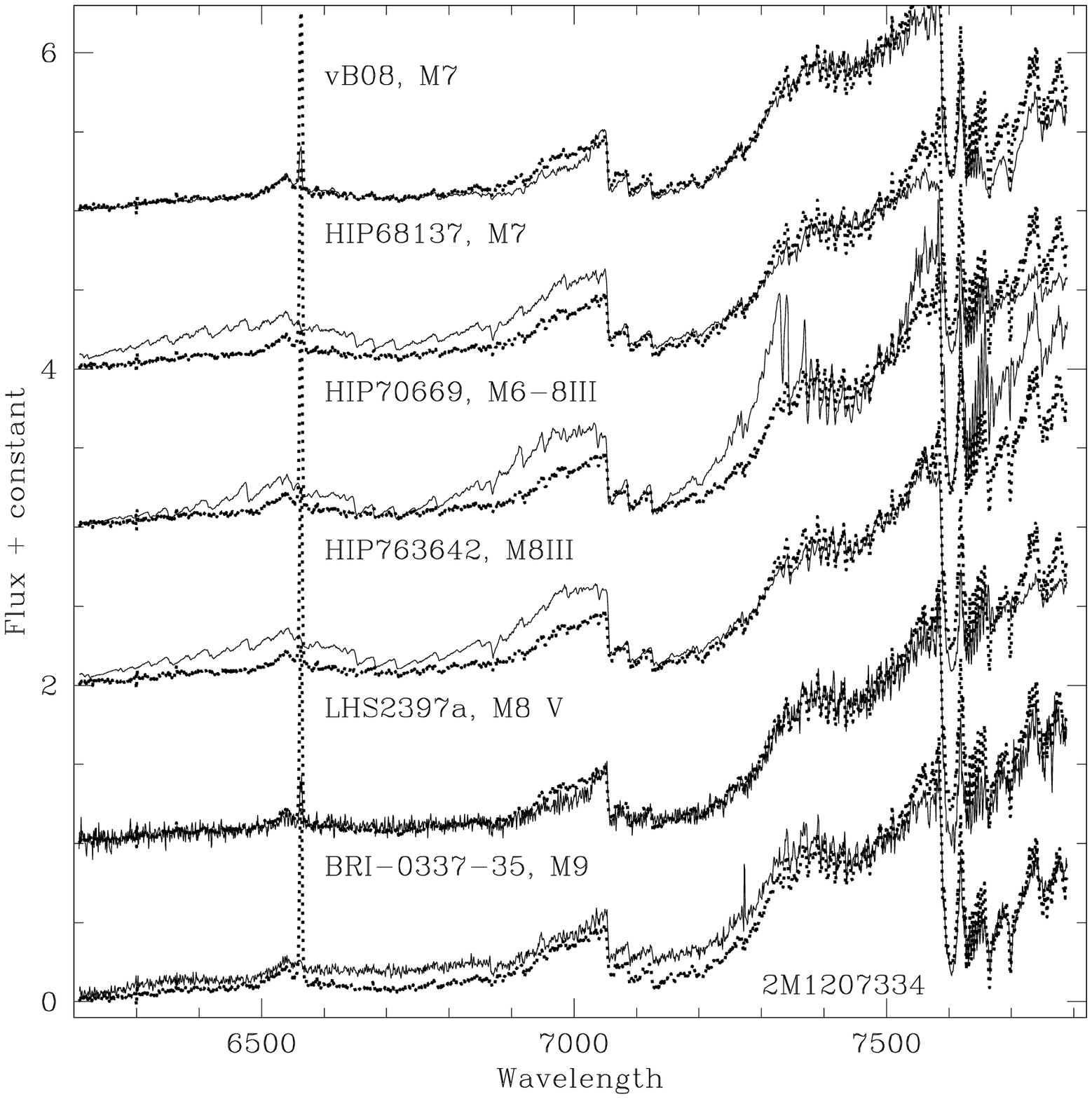}
 \caption{Spectral classification of 2M1139 and 2M1207.
We have normalized the spectra using the flux around 7450 \AA.
}
 \end{figure*}
%______________________________________________________________      
%%%%%%%%%%%%%%%%%%%%%
%%%%%%%%%%%%%%%%%%%%%

%-----------------------------------------------------------
    \begin{figure*}
    \centering
    \includegraphics[width=7.8cm]{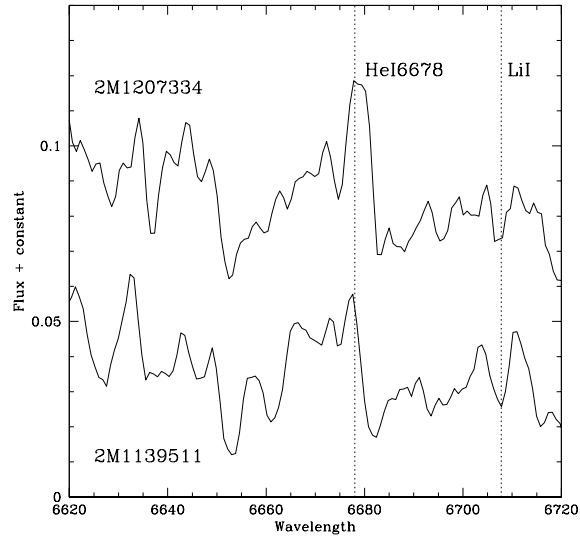}
 \caption{
 Spectra around the HeI6678 and LiI6708 \AA{} features.
}
 \end{figure*}
%______________________________________________________________      
%%%%%%%%%%%%%%%%%%%%%
%%%%%%%%%%%%%%%%%%%%%

%-----------------------------------------------------------
    \begin{figure*}
    \centering
    \includegraphics[width=7.8cm]{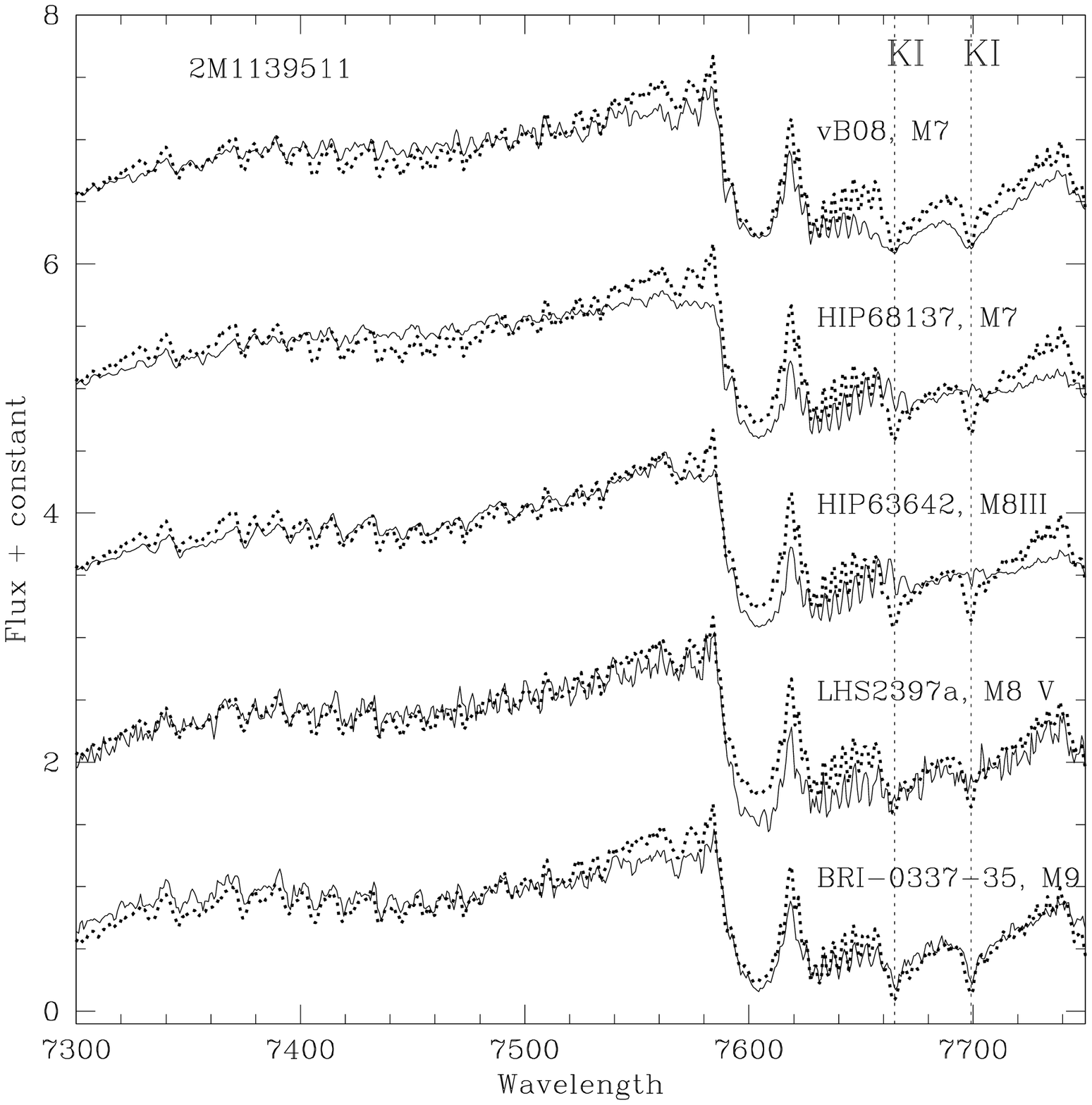}
    \includegraphics[width=7.8cm]{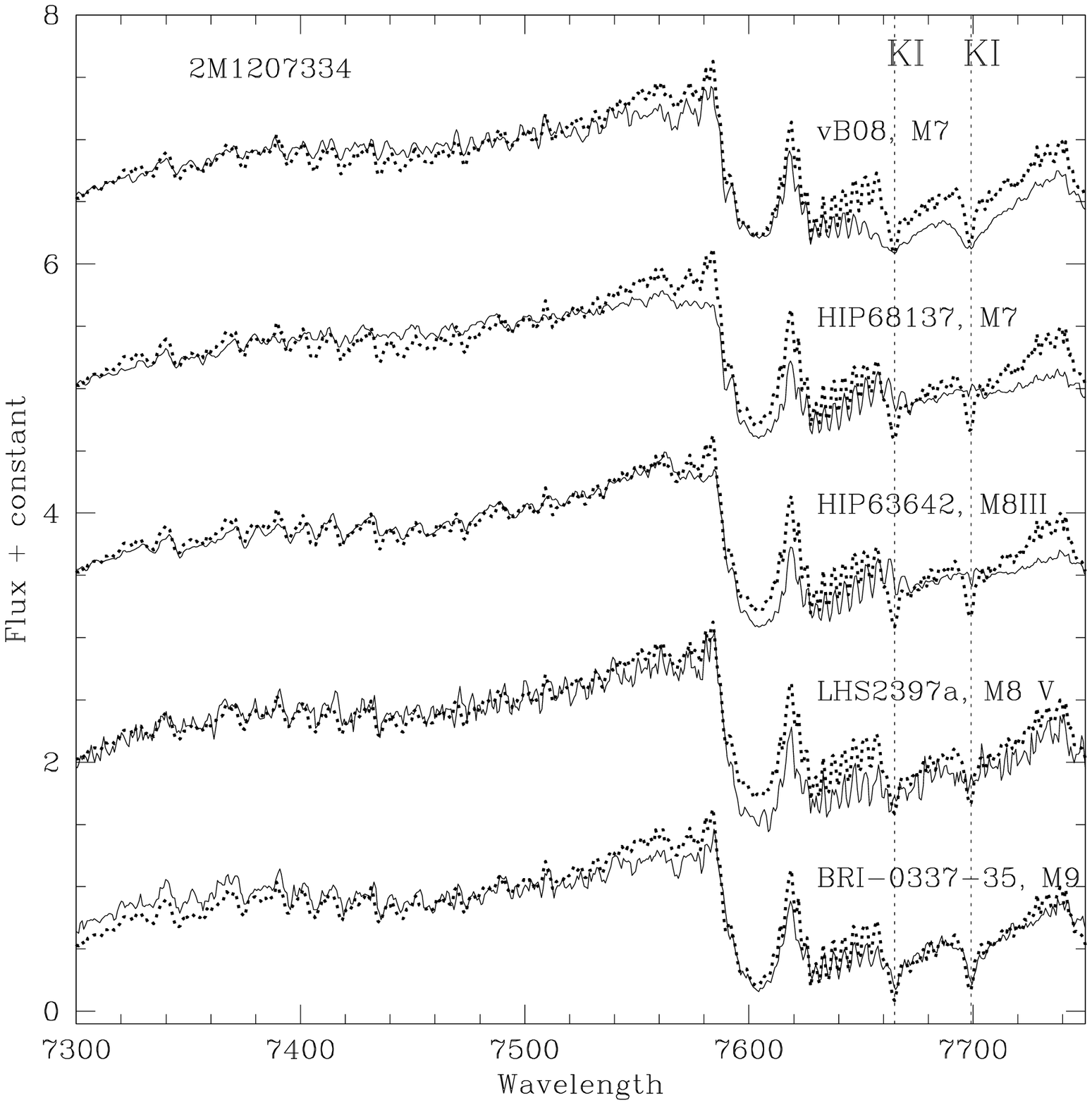}
 \caption{
Deriving the surface gravity by comparison with spectral templates.
}
 \end{figure*}
%______________________________________________________________      
%%%%%%%%%%%%%%%%%%%%%
%%%%%%%%%%%%%%%%%%%%%

%-----------------------------------------------------------
    \begin{figure*}
    \centering
    \includegraphics[width=7.8cm]{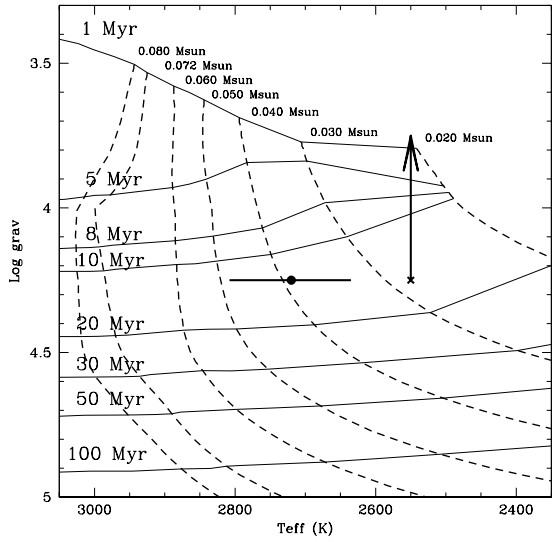}
 \caption{
The age for 2M1207, as derived from the surface gravity and the 
effective temperature (i.e., independent of the distance).
The solid circle represents  T$_{eff}$ derived from Luhman (1999) scale and 
  gravity  estimated by  comparison with spectral templates.
The arrow/cross indicates the location when the temperature derived by Mohanty (2006, priv.comm)
is used.
%The errorbars correspond to 2$\sigma$. 
The isochrones and  tracks correspond to models  by Baraffe et al. (1998).
}
 \end{figure*}
%______________________________________________________________      
%%%%%%%%%%%%%%%%%%%%%
%%%%%%%%%%%%%%%%%%%%%

%-----------------------------------------------------------
    \begin{figure*}
    \centering
    \includegraphics[width=7.8cm]{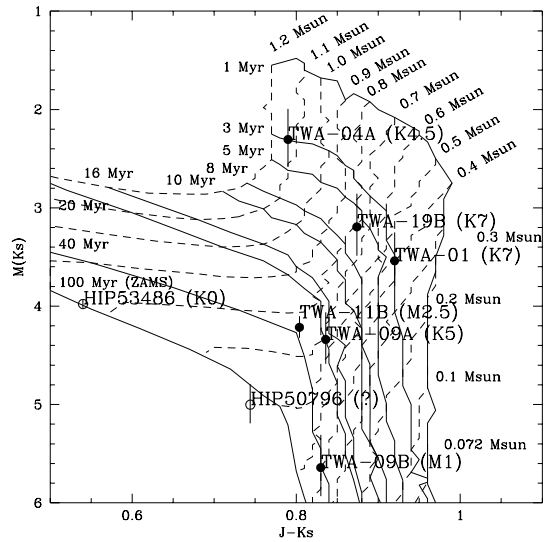}
 \caption{
Infrared Color-Color Diagram, showing models by Baraffe et al. (1998)
 and IR photometry from 2MASS.
}
 \end{figure*}
%______________________________________________________________      
%%%%%%%%%%%%%%%%%%%%%
%%%%%%%%%%%%%%%%%%%%%

%-----------------------------------------------------------
    \begin{figure*}
    \centering
    \includegraphics[width=5.8cm]{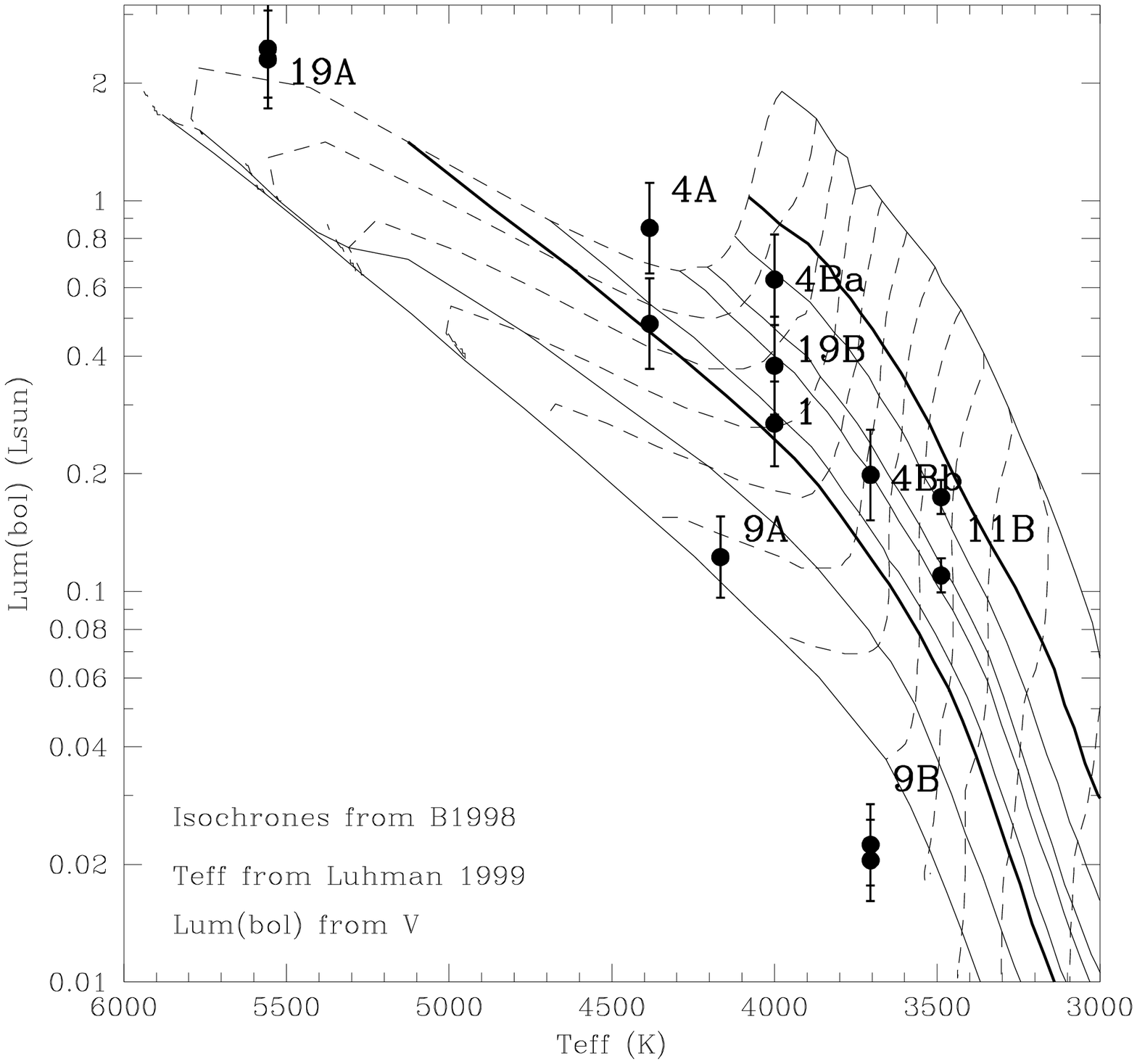}
    \includegraphics[width=5.8cm]{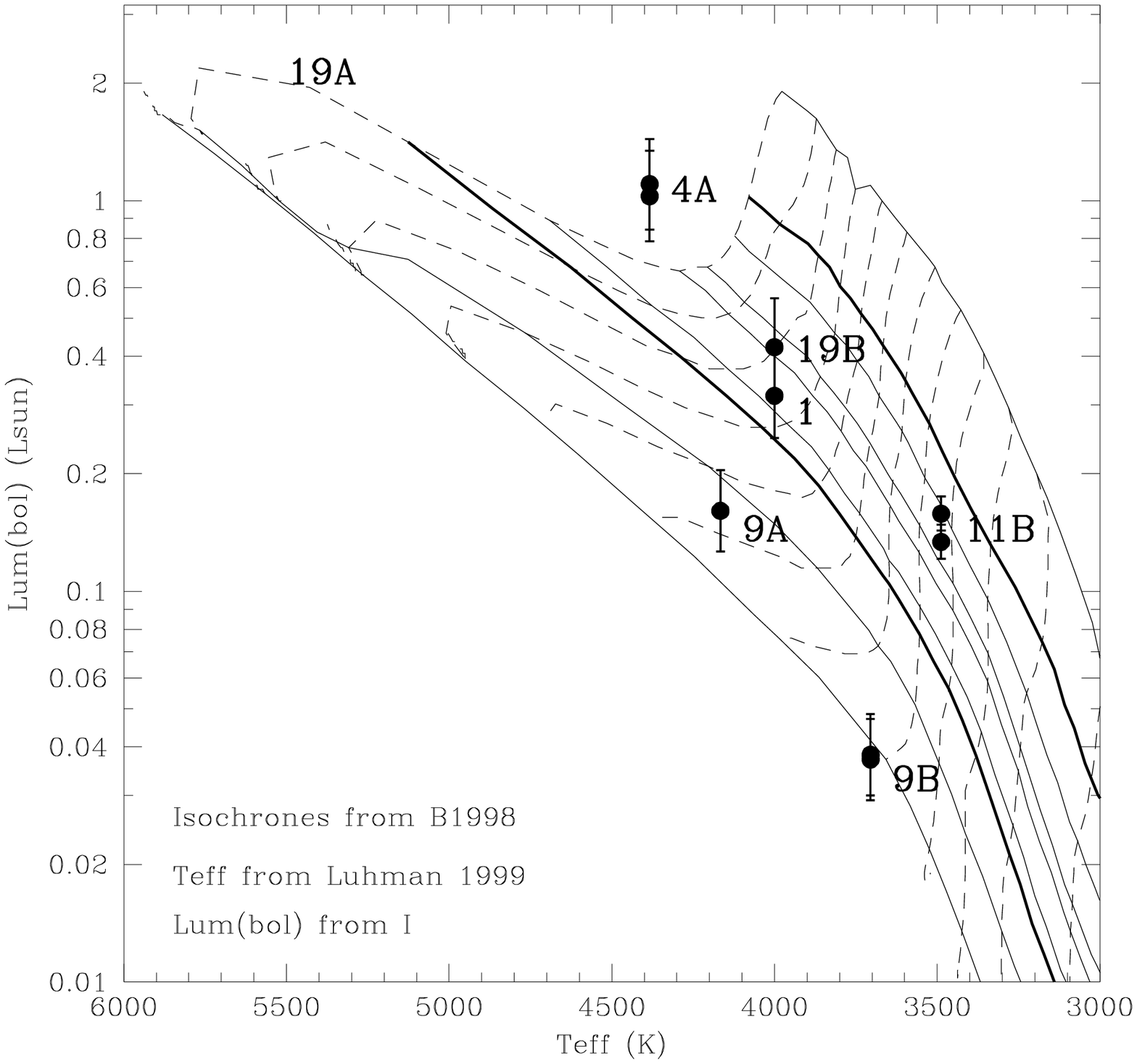}
    \includegraphics[width=5.8cm]{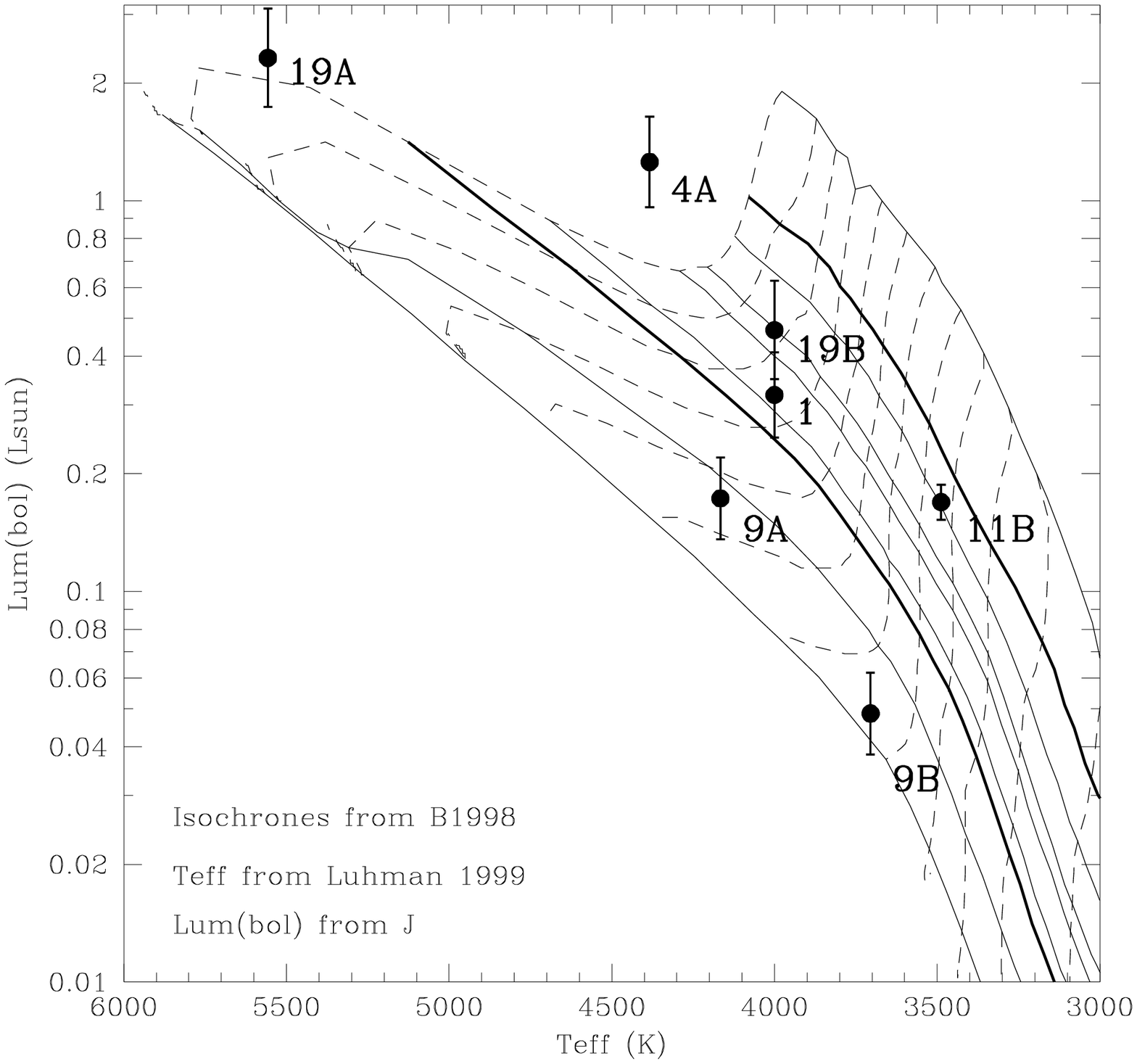}
    \includegraphics[width=5.8cm]{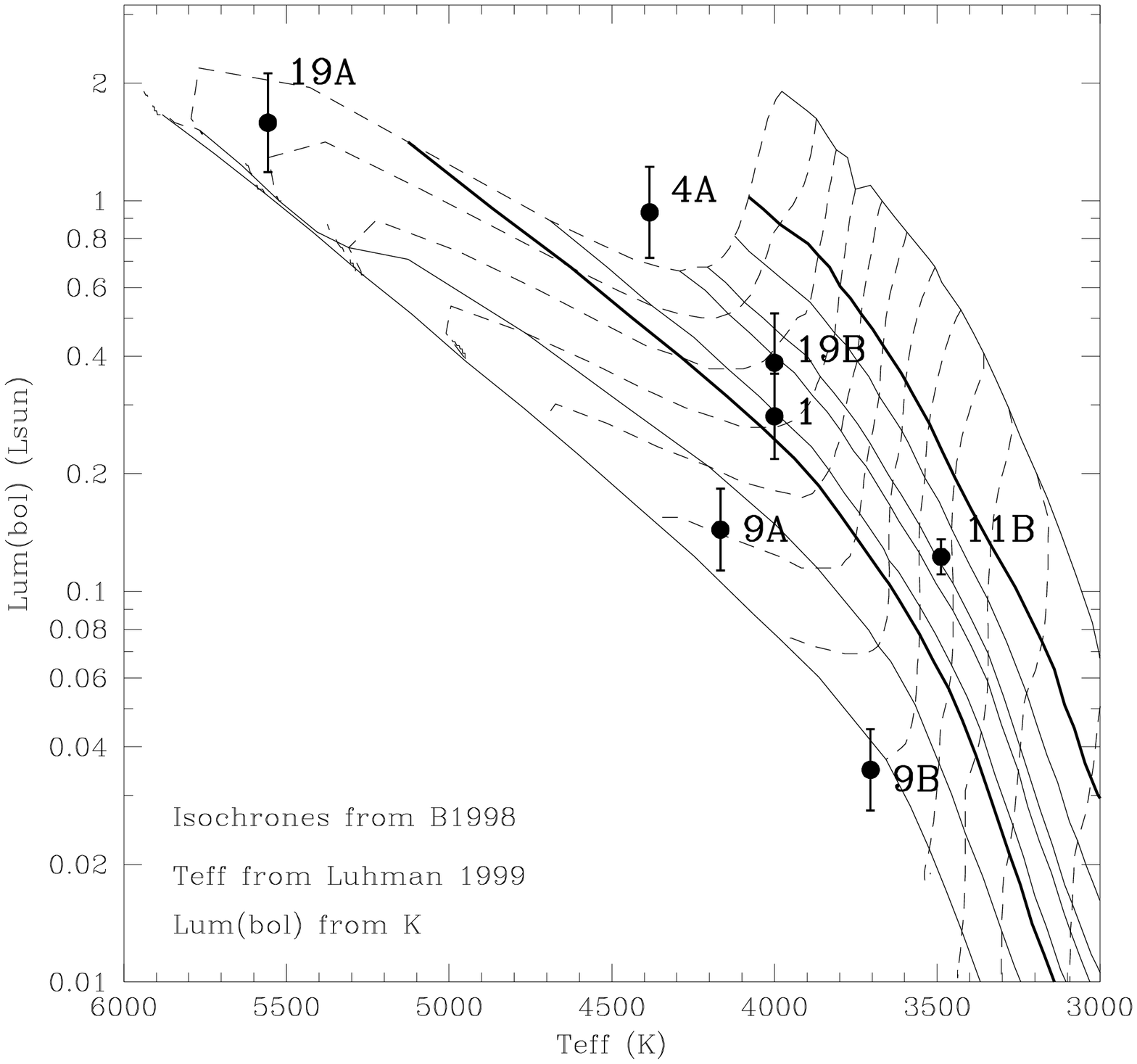}
 \caption{Bolometric luminosity --in logarithmic scale-- 
versus the effective temperature.
The luminosities have been derived from the filters $V$, $I$, $J$ \&
$K$  and the corresponding bolometric corrections (see text)
 in different panels. The evolutionary isochrones --solid lines-- 
and tracks --dashed lines-- correspond
to models by Baraffe et al. (1998), whereas the effective temperatures 
were derived with the scale by Luhman (1999) for intermediate gravity.
Errors in luminosities include the error in the distance and the photometric
error when available.}
 \end{figure*}
%______________________________________________________________      
%%%%%%%%%%%%%%%%%%%%%
%%%%%%%%%%%%%%%%%%%%%

%-----------------------------------------------------------
    \begin{figure*}
    \centering
    \includegraphics[width=6.8cm]{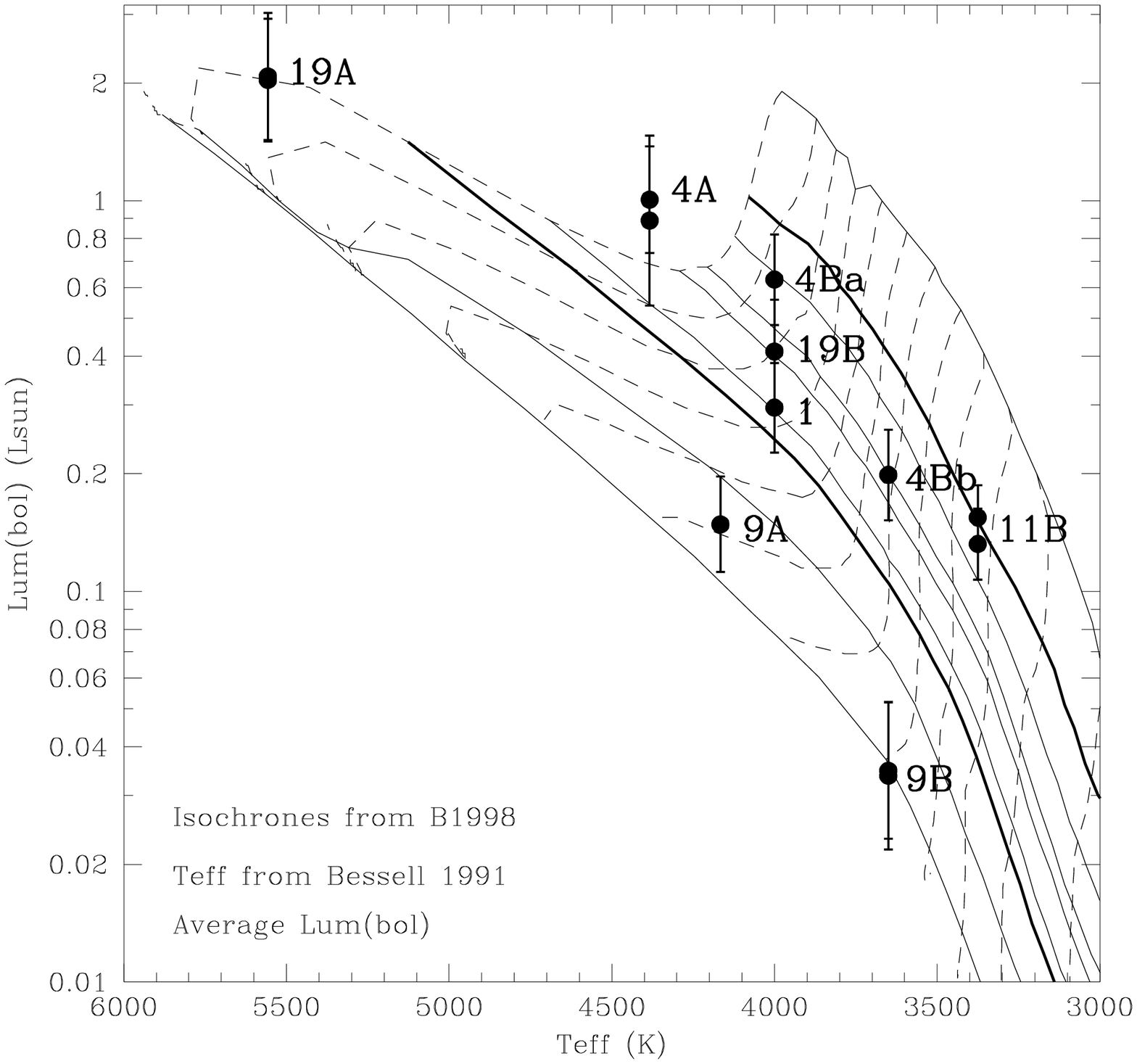}
    \includegraphics[width=6.8cm]{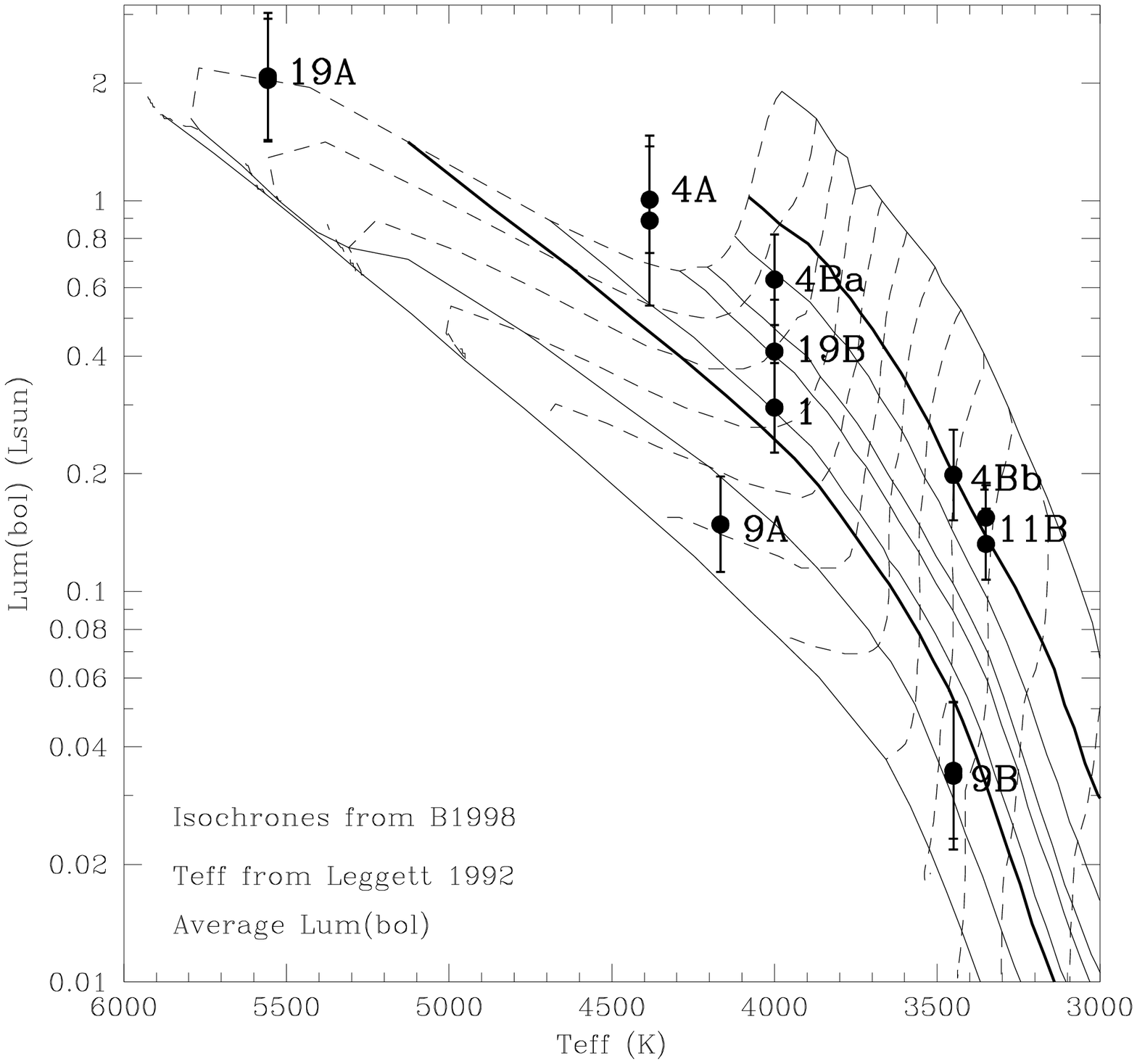}
    \includegraphics[width=6.8cm]{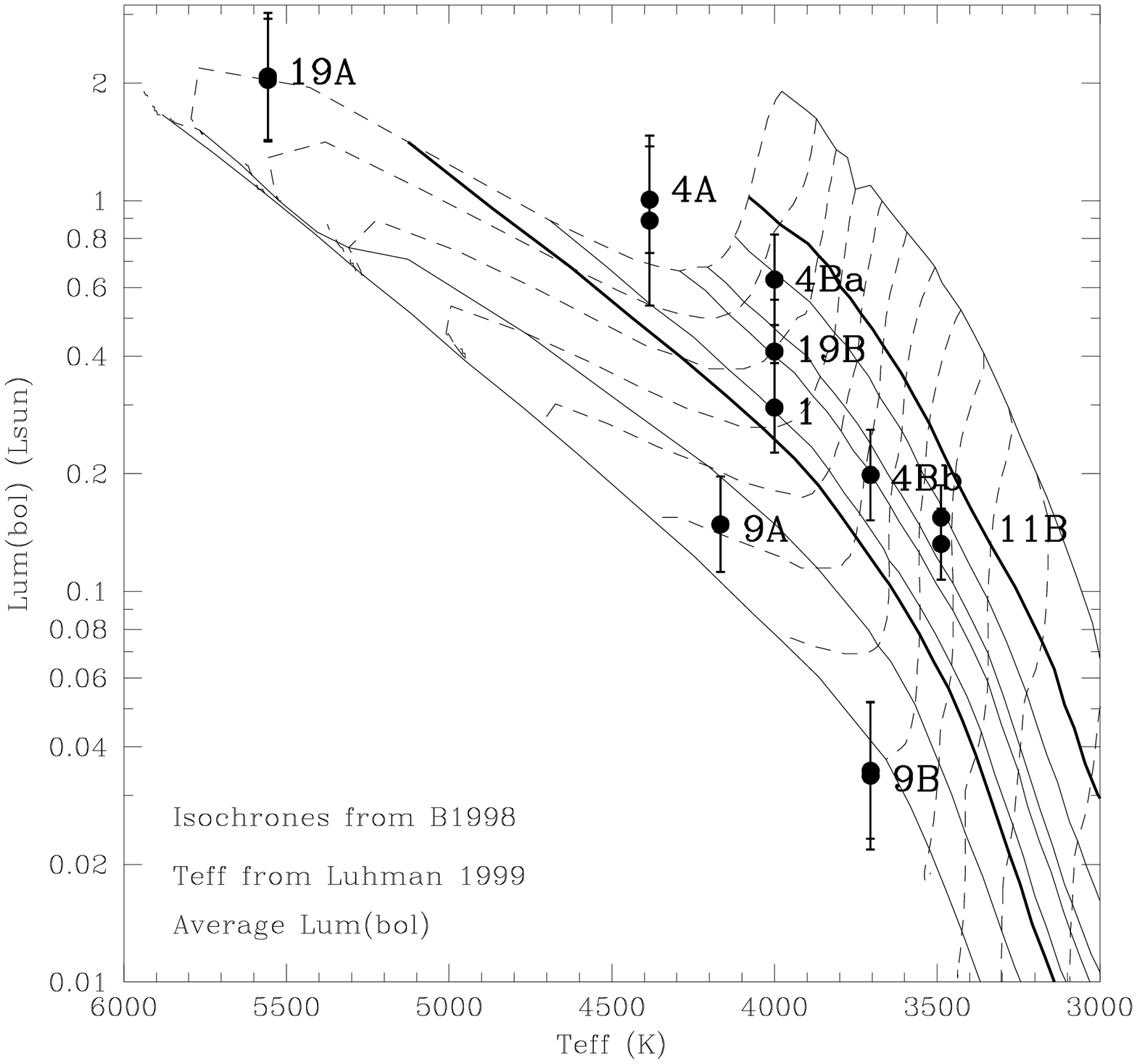}
 \caption{Bolometric luminosity --in logarithmic scale-- 
versus the effective temperature.
The last values were computed using the temperature scales
in the case of M spectral types
of Bessell (1991), Leggett et al. (1992) and  Luhman (1999).
For warmer stars,  Bessell (1979) was used in all three cases.
The bolometric luminosities correspond to average values.
Errors in luminosities include the error in the distance and the 
dispersion of the individual values used when computing the  average.
 The evolutionary isochrones --solid lines--  and
 tracks --dashed lines-- correspond
to models by Baraffe et al. (1998).}
 \end{figure*}
%______________________________________________________________      
%%%%%%%%%%%%%%%%%%%%%
%%%%%%%%%%%%%%%%%%%%%

%-----------------------------------------------------------
    \begin{figure*}
    \centering
    \includegraphics[width=6.8cm]{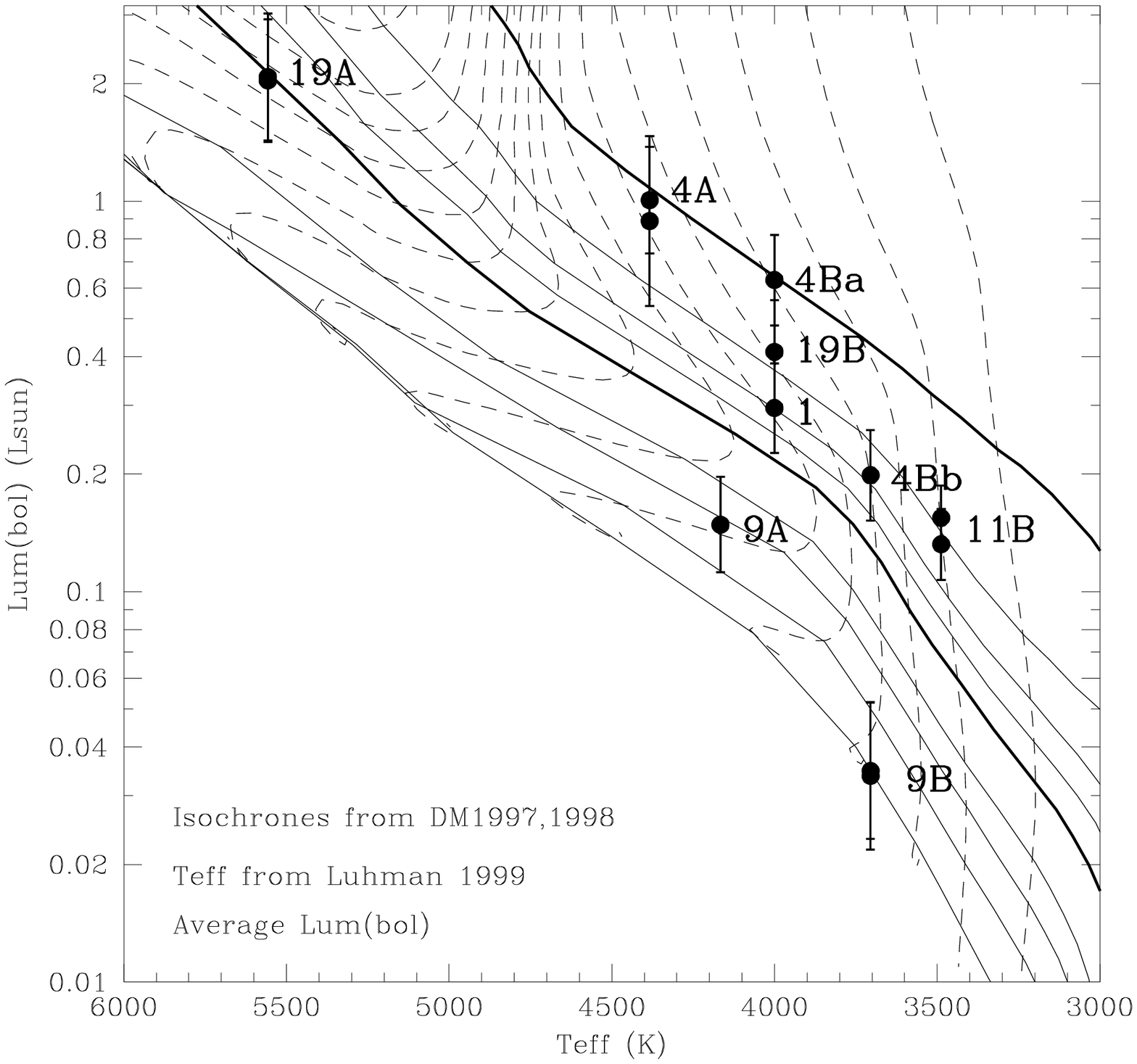}
    \includegraphics[width=6.8cm]{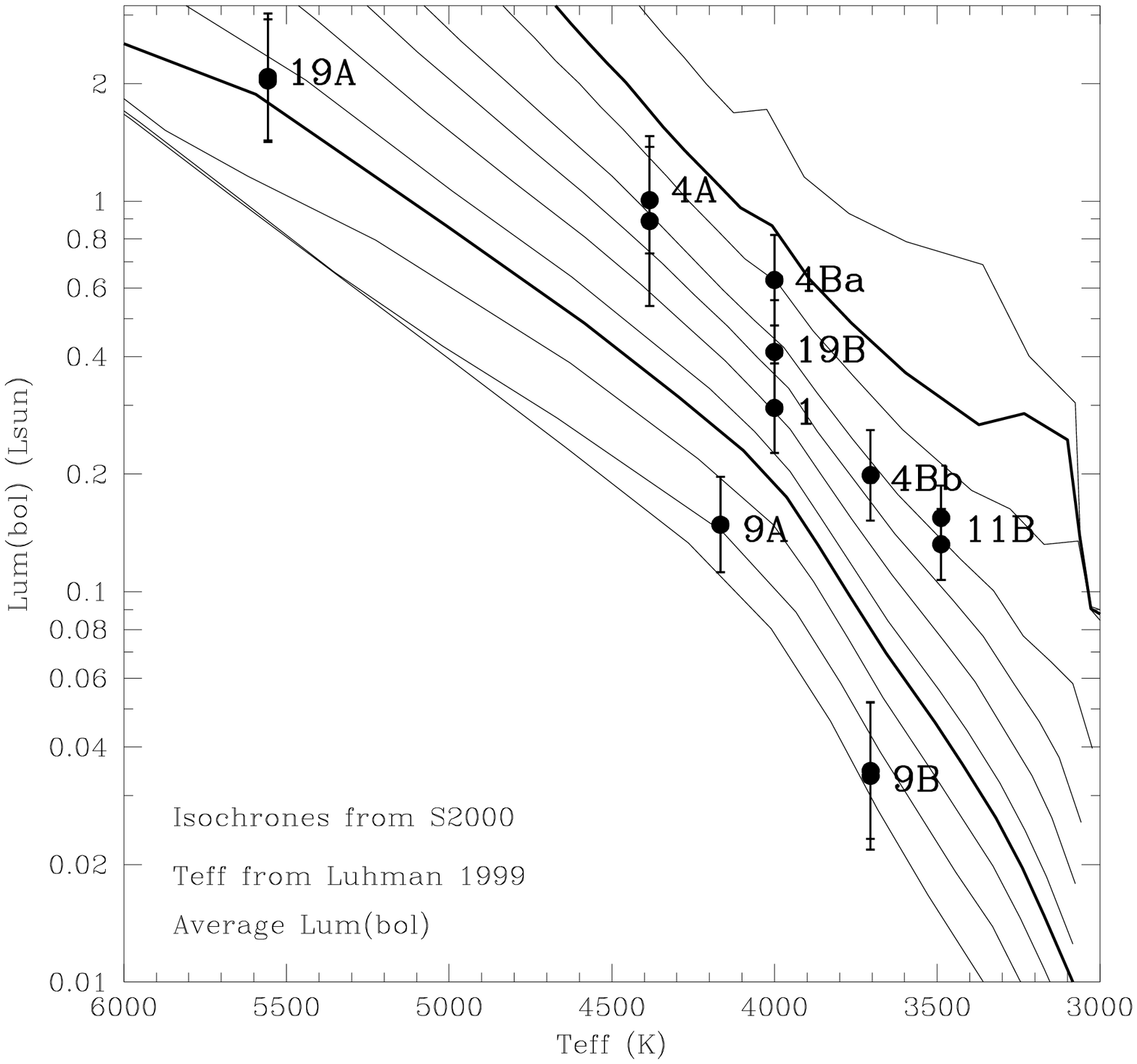}
    \includegraphics[width=6.8cm]{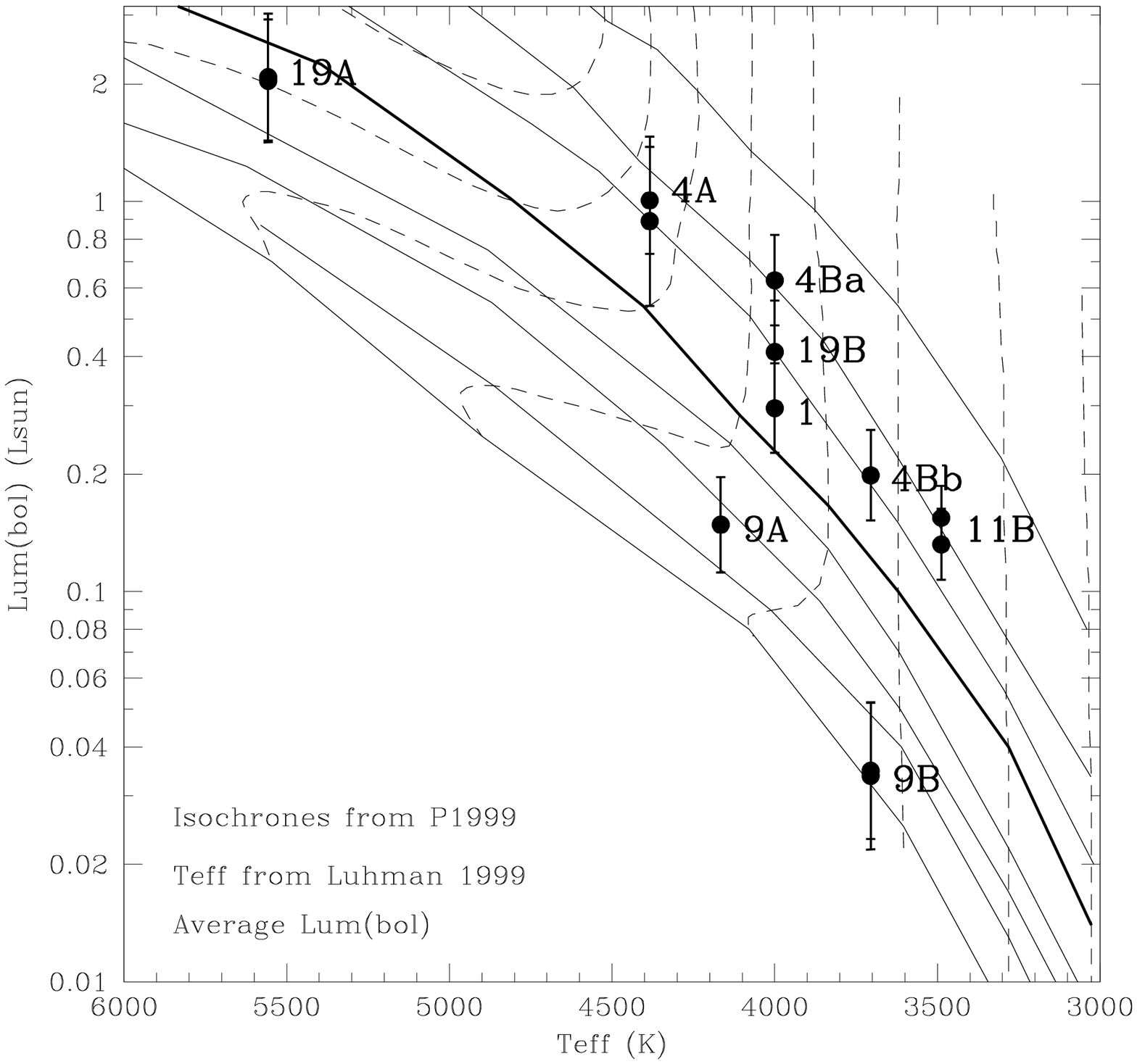}
 \caption{Bolometric luminosity --in logarithmic scale-- 
versus the effective temperature.
 The evolutionary isochrones and tracks correspond
to models  by D'Antona \& Mazzitelli (1997, 1998) 
 Siess et al. (2000) and Palla \& Stahl (1999).
Models by Baraffe et al. (1998) are displayed in the previous figure.
The bolometric luminosities correspond to average values.
Errors in luminosities include the error in the distance and the 
dispersion of the individual values used when computing the  average.
The effective temperatures 
were derived with the scale by Luhman (1999) for intermediate gravity.}
 \end{figure*}
%______________________________________________________________      
%%%%%%%%%%%%%%%%%%%%%
%%%%%%%%%%%%%%%%%%%%%

%-----------------------------------------------------------
    \begin{figure*}
    \centering
    \includegraphics[width=7.8cm]{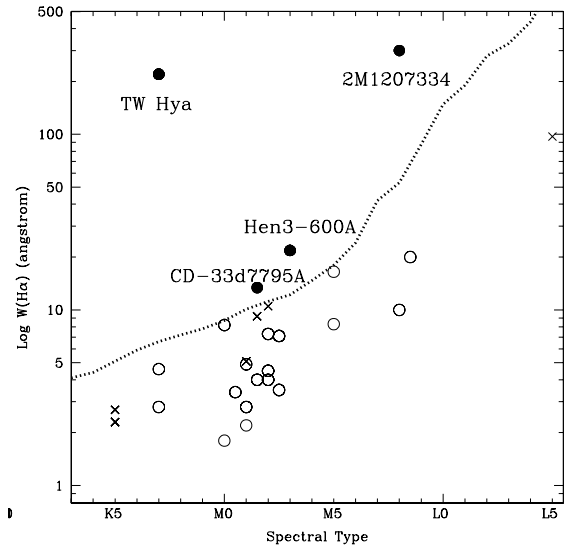}
 \caption{
H$\alpha$ equivalent width --in logarithm-- versus the spectral type.
The  dotted line is the accretion criterion proposed by Barrado y Navascu\'es 
\& Mart\'{\i}n (2003).
Classical and weak-line/post TTauri stars or substellar analogs are displayed with solid 
or open circles, respectively. Probable non-members  are included as crosses.
}
 \end{figure*}
%______________________________________________________________      
%%%%%%%%%%%%%%%%%%%%%
%%%%%%%%%%%%%%%%%%%%%

%-----------------------------------------------------------
    \begin{figure*}
    \centering
    \includegraphics[width=7.8cm]{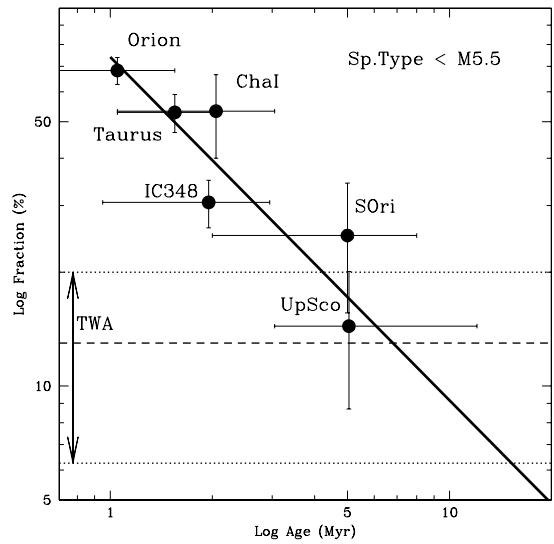}
 \caption{
Fraction of stars classified as  CTT stars  (based on the H$\alpha$ equivalent width)
for different star forming regions and young clusters.
The dashed line corresponds to the average value for TWA, whereas the two dotted
lines represent the maximum and minimum value. The clear trend with the age (solid line)
indicates that  TWA is about 7 Myr old.
}
 \end{figure*}
%______________________________________________________________      
%%%%%%%%%%%%%%%%%%%%%
%%%%%%%%%%%%%%%%%%%%%

%-----------------------------------------------------------
    \begin{figure*}
    \centering
    \includegraphics[width=7.8cm]{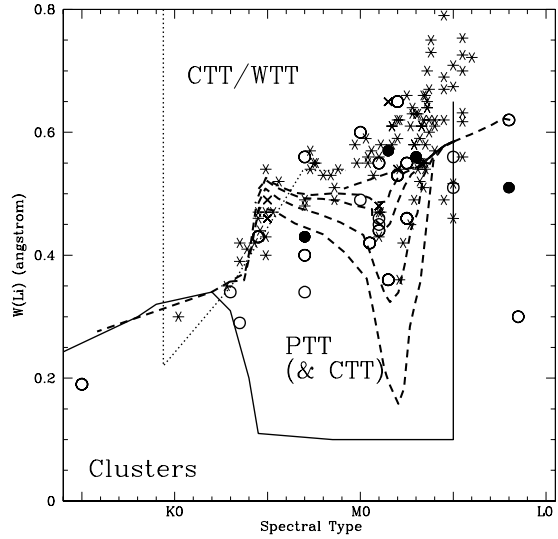}
 \caption{
Lithium equivalent with versus the spectral type.
The solid line corresponds to the upper envelope of the values
measured in young open clusters such as IC2391, IC2602, the Pleiades and M35.
The long-dashed line delimits the areas for weak-line and post-TTauri
stars (adapted from Mart\'{\i}n 1997 and Mart\'{\i}n \& Magazz\`u 1999).
Thick, short-dashed  lines correspond to Baraffe et al. (1998) lithium depletion
isochrones (using curves of growth from Zapatero Osorio et al$.$ 2002 and the 
temperature scale by Luhman 1999). We have represented values corresponding to
1, 8, 10, 15 and 20 Myr isochrones. The first one has no lithium depletion
--A(Li)=3.1.
TWA members are displayed as solid circles (CTT), open circles (WTT/PTT) and
probable non-members (crosses).
The asterisks correspond to members of the  Lambda Orionis cluster 
(after Dolan \& Mathieu 1999, 2001;
 and Barrado y Navascu\'es et al. 2004). This cluster is about 5 Myr, with a 
maximum age of 8 Myr. Based on the lithium equivalent width,  the TW Hydrae association 
is older than the Lambda Orionis cluster. 
}
 \end{figure*}
%______________________________________________________________      
%%%%%%%%%%%%%%%%%%%%%
%%%%%%%%%%%%%%%%%%%%%

%-----------------------------------------------------------
    \begin{figure*}
    \centering
    \includegraphics[width=4.0cm]{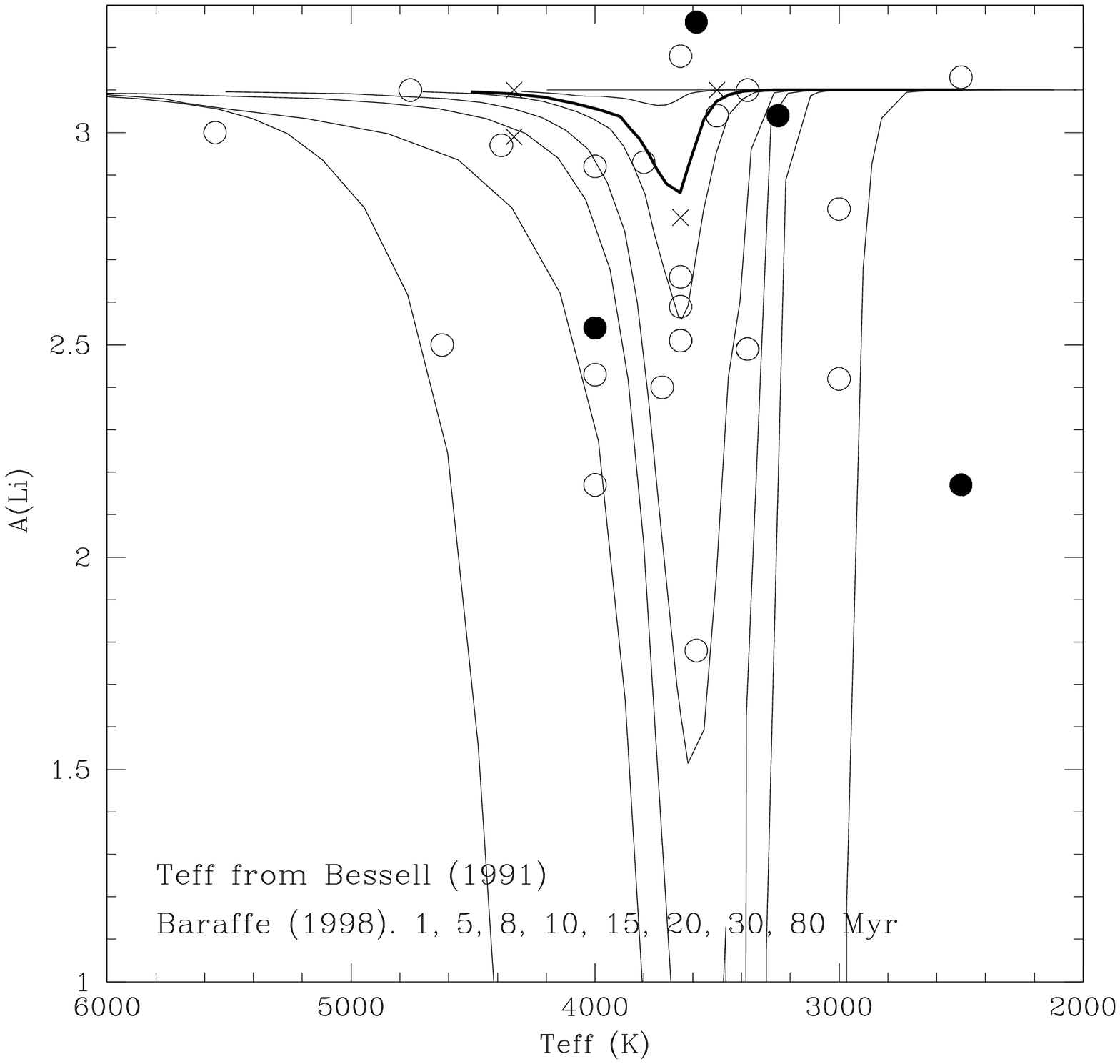}
    \includegraphics[width=4.0cm]{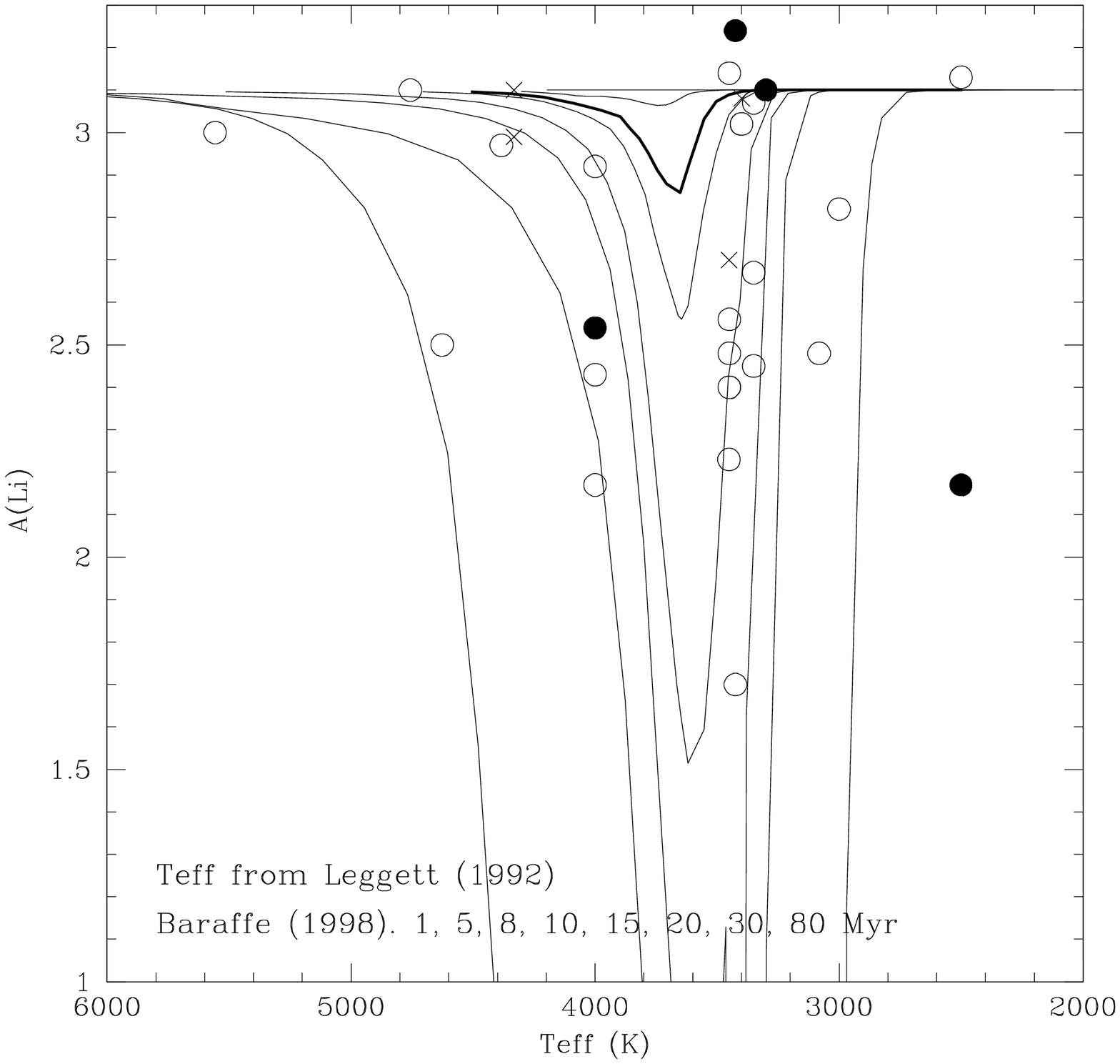}
    \includegraphics[width=4.0cm]{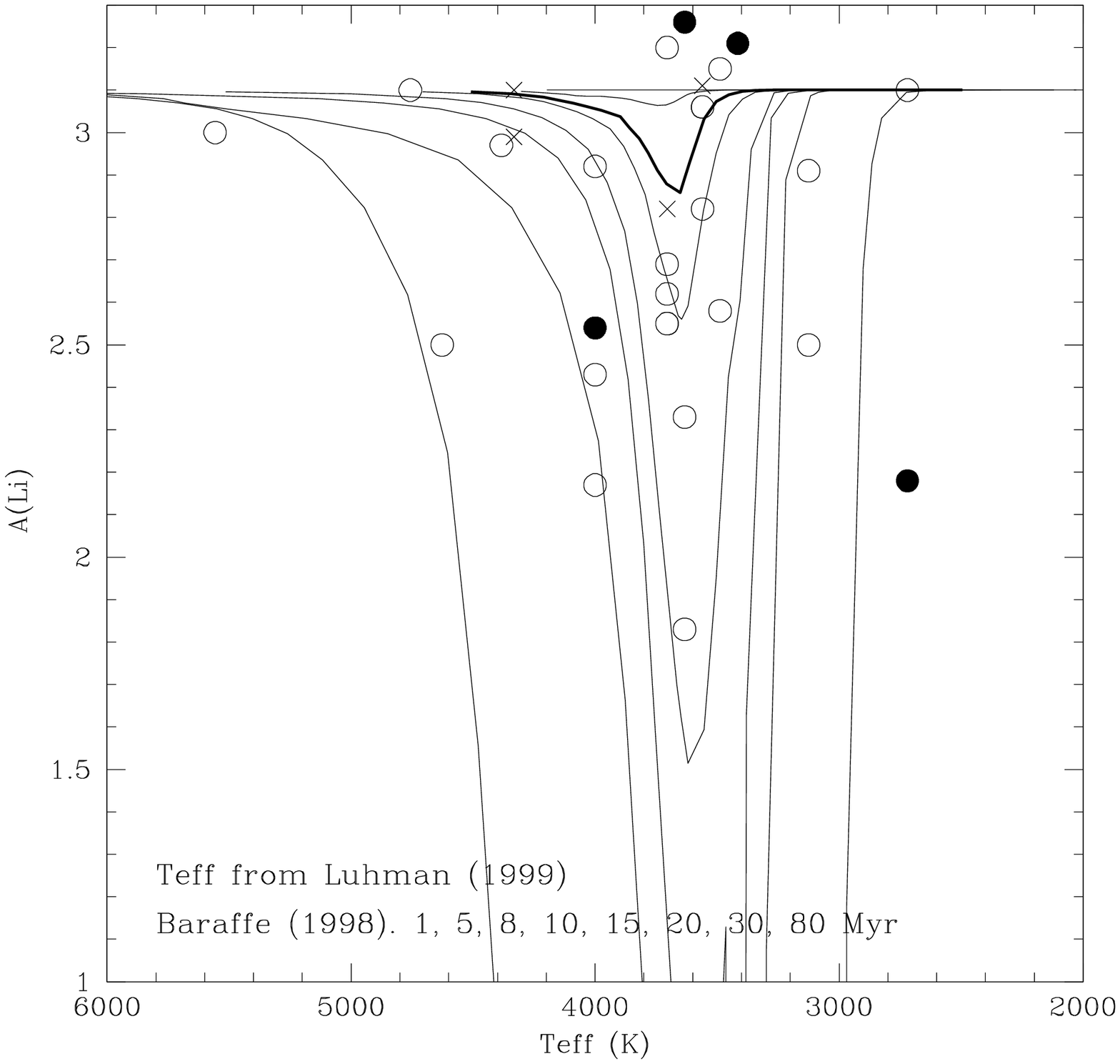}

    \includegraphics[width=4.0cm]{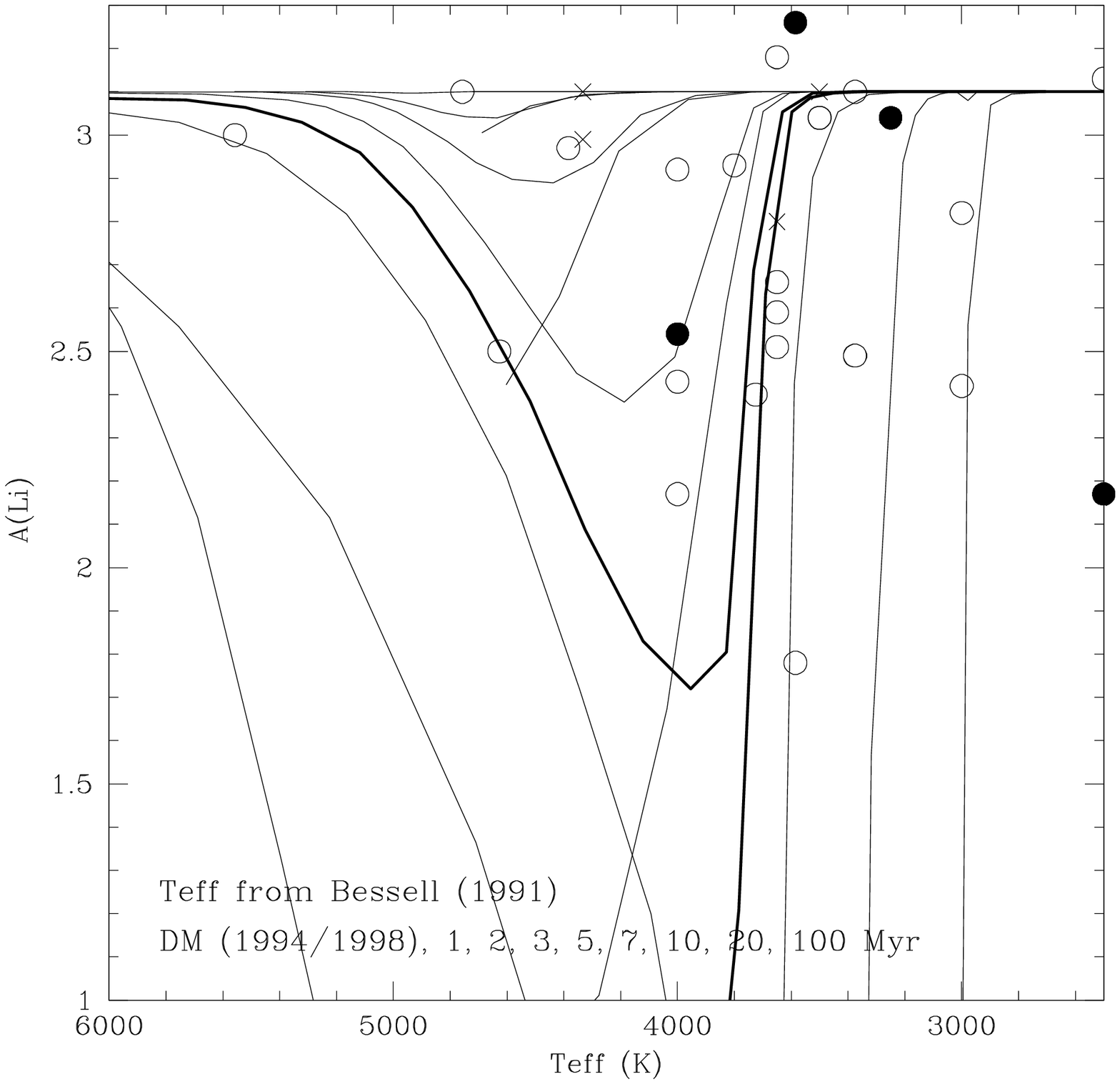}
    \includegraphics[width=4.0cm]{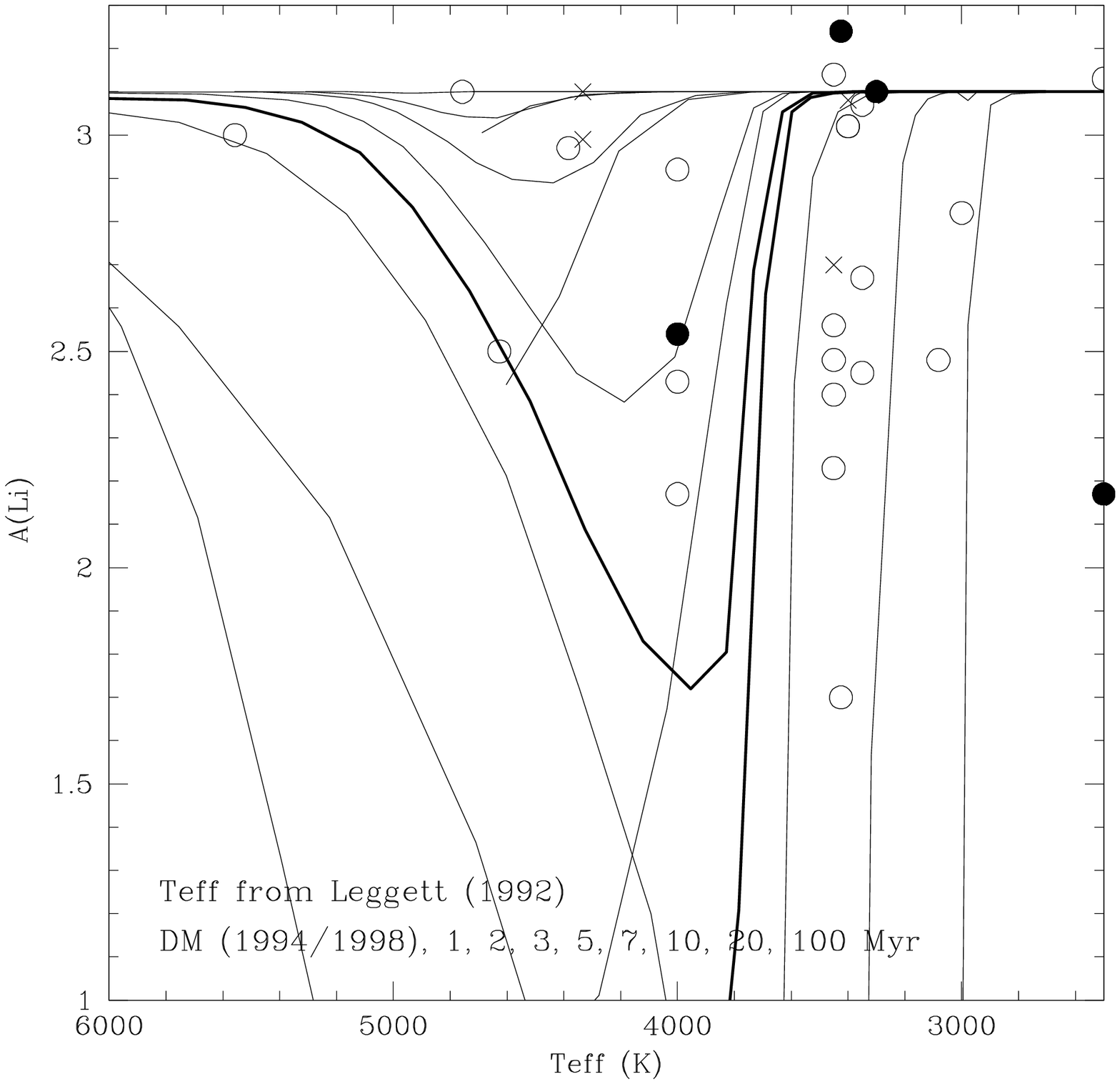}
    \includegraphics[width=4.0cm]{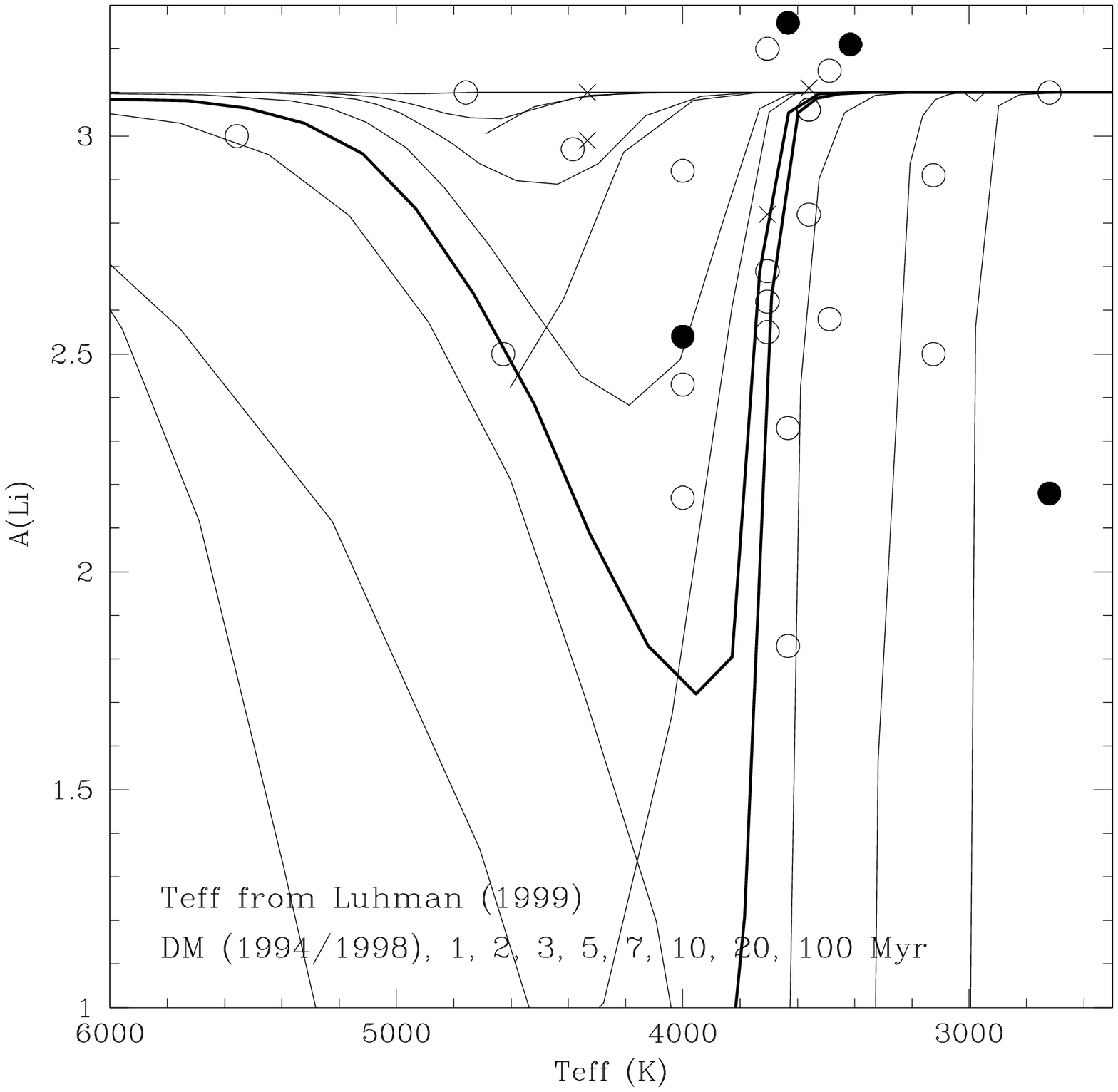}

    \includegraphics[width=4.0cm]{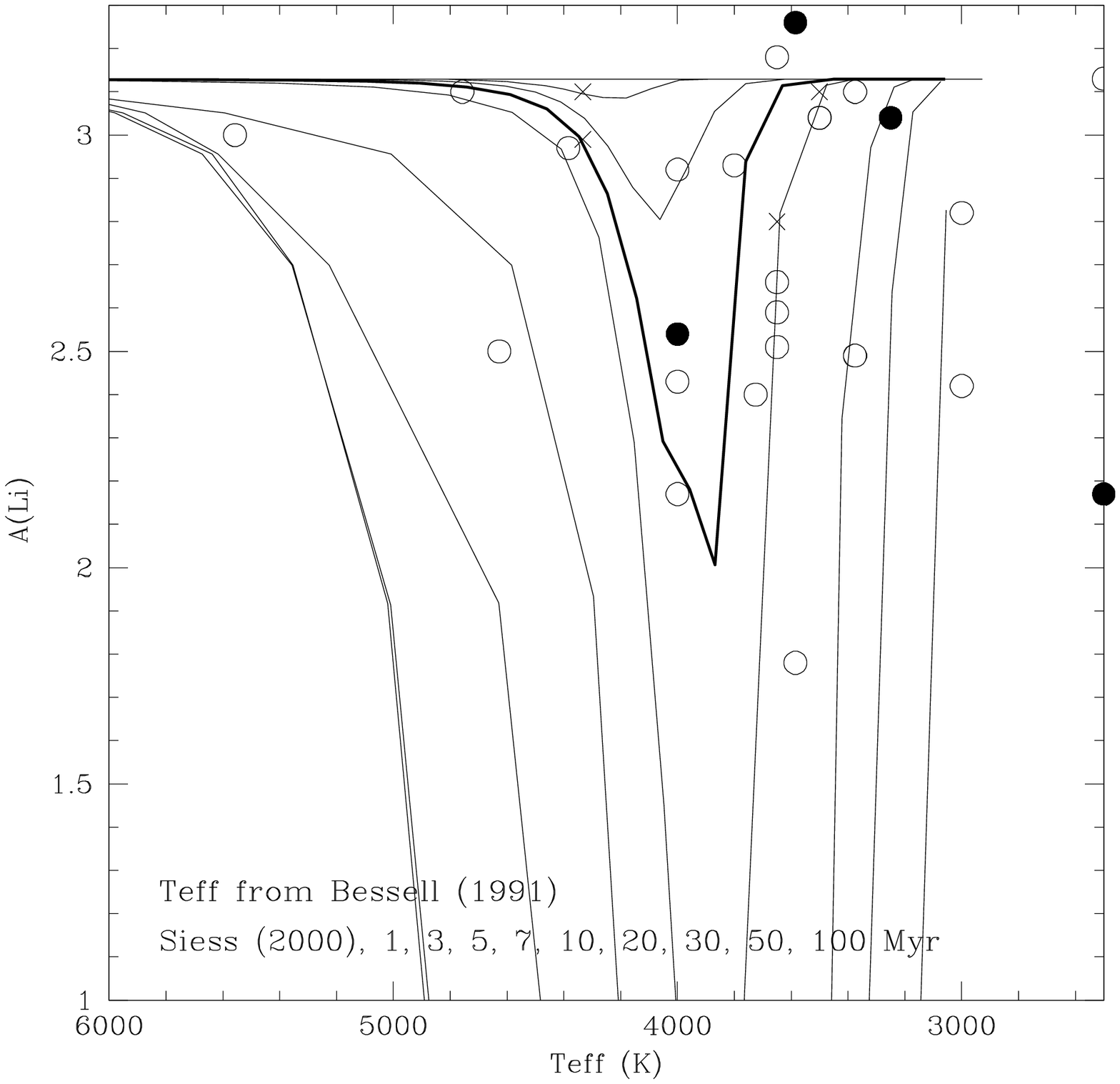}
    \includegraphics[width=4.0cm]{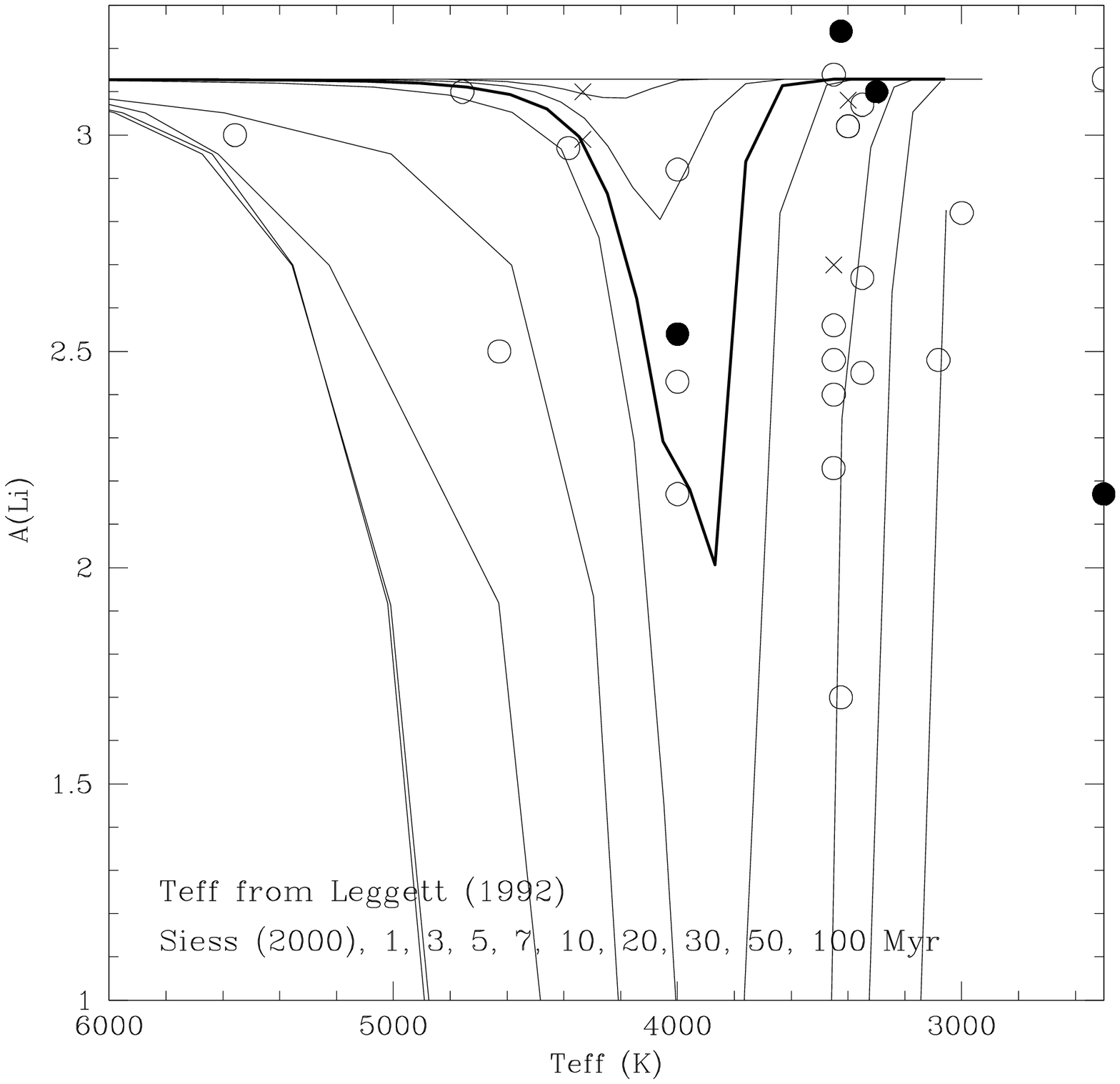}
    \includegraphics[width=4.0cm]{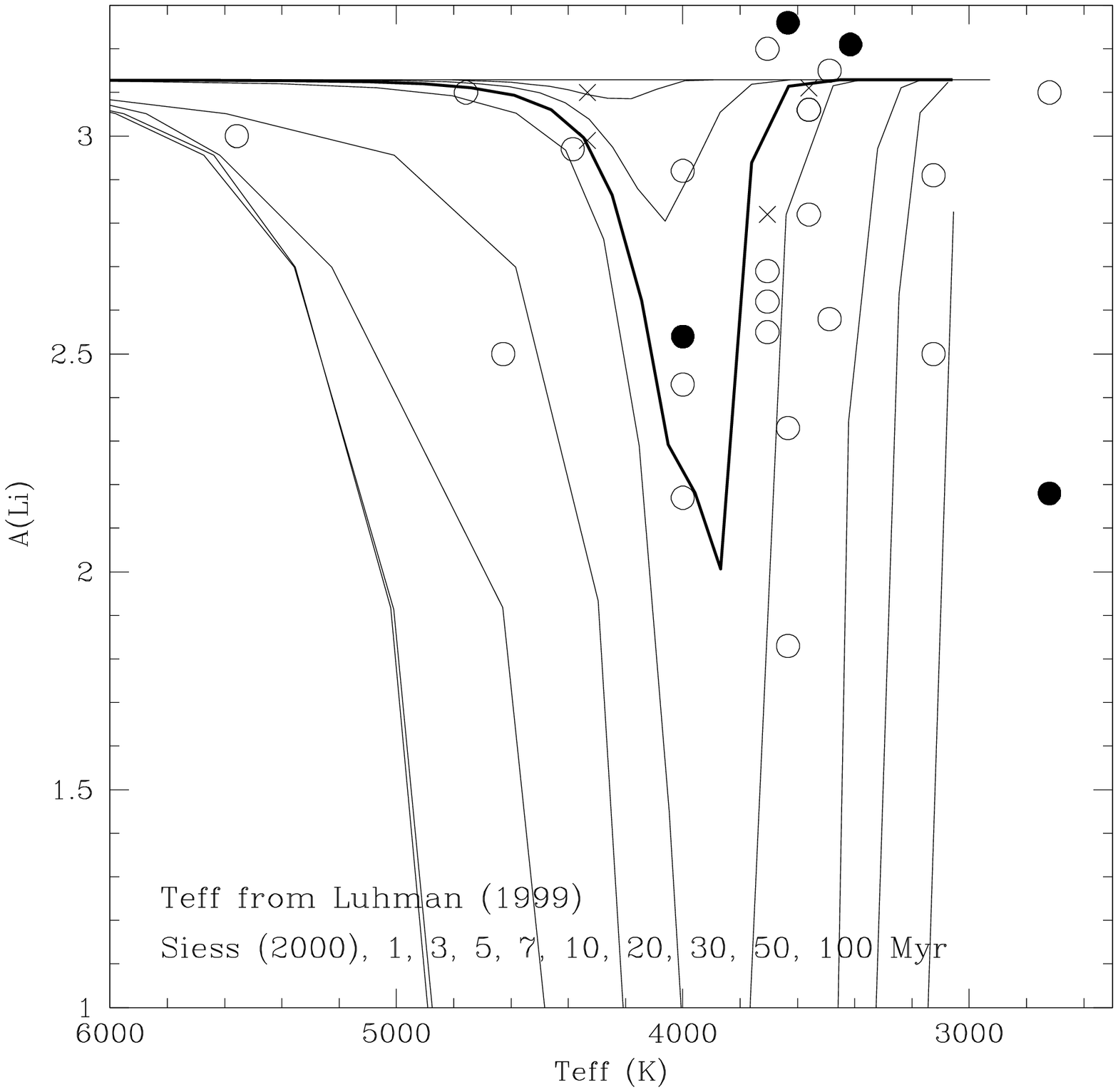}
 \caption{
Lithium abundance versus the spectral temperature for different
models and temperature scales.
TWA members are displayed as solid circles (CTT), open circles (WTT/PTT) and
probable non-members (crosses).
The 7/8 Myr Li depletion isochrone is highlighted in most of the 
figures.
}
 \end{figure*}
%______________________________________________________________      
%%%%%%%%%%%%%%%%%%%%%
%%%%%%%%%%%%%%%%%%%%%

\setcounter{table}{0}
\begin{table*}[ht]
\caption[]{Data for members of the  TW Hydrae association (ONLY IN THE ELECTRONIC VERSION)}
\small
\begin{tabular}{lccccccclcl}
TWA       &Sp.Type&  ref.&  V    & I      &Parallax&W(Halpha)&ref  &  W(Li)&  ref &  Name  \\
\hline
TWA-01    &  K7e  & (02) & 10.92 &  9.18  & 17.72 & 220    &(08)   & 0.39  &(08)  & TWHya                   \\% (10)  &      (14) & 
''        &  K8Ve & (10) & --    &  --    & --    & 265    &(02)   & 0.46  &(02)  & --                      \\% --    &     --   & 
''        &  --   & --   & --    &  --    & --    & 300.   &(10)   & 0.440 &(10)  & --                      \\% --    &     --   & 
''        &  --   & --   & --    &  --    & --    & 150.0  &(07)   & 0.530 &(07)  & --                      \\% --    &     --   & 
''        &  --   & --   & --    &  --    & --    & 238.0  &(11)   & 0.426 &(18)  & --                      \\% --    &     --   & 
''        &  --   & --   & --    &  --    & --    & 165.0  &(11)   & 0.37  &(06)  & --                      \\% --    &     --   & 
''        &  --   & --   & --    &  --    & --    & 225    &(06)   & --    & --   & --                      \\% --    &     --   & 
''        &  --   & --   & --    &  --    & --    &71.4-101&(01)   & --    & --   & --                      \\% --    &    --   & 
''        &  --   & --   & --    &  --    & --    & 172.5  &(23)   & --    & --   & --                      \\% --    &    --   & 
                                                                                                                                         
TWA-02A   &  M0.5 & (08) & 11.60 &  9.20  & --    &   1.89 &(08)   & 0.494 &(18)  & CD-29d8887A             \\% (16)  &     (14) & 
''        &  M2Ve & (10) & --    &  --    & --    &   2    &(02)   & 0.49  &(08)  & --                      \\% --    &             --   & 
''        &  M0   & (02) & --    &  --    & --    &   2.1  &(10)   & 0.560 &(10)  & --                      \\% --    &             --   & 
''        &  --   & --   & --    &  --    & --    &  --    & --    & 0.52  &(02)  & --                      \\% --    &      --   & 
''        &  --   & --   & --    &  --    & --    &   1.8  &(23)   & --    & --   & --                      \\% --    &    --   & 
TWA-02B   &  M2   & (08) & 13.50 & 10.70  & --    &   2.28 &(06)   & 0.42  &(06)  & CD-29d8887B             \\% (16)  &     (14) & 
                                                                                                                                         
TWA-03A   &  M3e  & (02) & 12.00 &  9.60  & --    &  21.8  &(08)   & 0.53  &(08)  & Hen3-600A               \\% (16)  &     (14) & 
''        &  M3   & (08) & 12.04 &  9.10  & --    &  10.   &(10)   & 0.610 &(10)  & --                      \\% (10)  &     --   & 
''        &  M4Ve & (10) & --    &  9.08  & --    &  20.   &(02)   & 0.55  &(02)  & --                      \\% (02)  &                    --   & 
''        &  --   & --   & --    &  --    & --    &  26.0  &(11)   & 0.563 &(18)  & --                      \\% --    &                    --   & 
''        &  --   & --   & --    &  --    & --    &  27.0  &(11)   & --    & --   & --                      \\% --    &                    --   & 
''        &  --   & --   & --    &  --    & --    &  22.6  &(23)   & --    & --   & --                      \\% --    &    --   & 
                                                                                                                                         
TWA-03B   &  M3.5 & (08) & 13.70 &  8.86  & --    &   7.14 &(08)   & --    & --   & Hen3-600B               \\% (02)  &     (14) & 
''        &  M3   & (02) & --    & 10.10  & --    &   8.   &(02)   & 0.54  &(08)  & --                      \\% (16)  &                    --   & 
''        &  --   & --   & --    &  --    & --    &   6.7  &(11)   & 0.50  &(02)  & --                      \\% --    &                    --   & 
''        &  --   & --   & --    &  --    & --    &   6.1  &(23)   & --    & --   & --                      \\% --    &    --   & 
                                                                                                                                                     
TWA-04A   &  K4-K5& (05) &  8.89 &  7.50  & 21.43 &   0.   &(14)   & 0.360 &(14)  & HD98800A                \\% (16)  &     (14) & 
''        &  K7V  & (10) &  9.41 &  7.39  & --    &   0.0  &(11)   & 0.425 &(05)  & --                      \\% (10)  &            --  &        
''        &  --   &  --  &  --   &  --    & --    &   0.   &(10)   & 0.430 &(10)  & --                      \\%   --  &                     --  & 
                                                                                                                                         
TWA-04Ba  &  K7   & (05) &  9.71 & --     & 21.43 &  --    & --    & 0.335 &(05)  & HD98800Ba               \\%  --   &     (14) & 
                                                                                                                                         
TWA-04Bb  &  M1   & (05) & 11.47 & --     & 21.43 &  --    & --    & 0.450 &(05)  & HD98800Bb               \\%  --   &(14) &
                                                                                                                                         
TWA-05Aab &  M1.5 & (08) & 11.54 &  9.01  & --    &  13.4  &(08)   & 0.57  &(08)  & CD-33d7795A             \\% (10)  &     (14) & 
''        &  M3Ve & (10) & 11.72 &  9.10  & --    &  10.   &(10)   & 0.540 &(10)  & --                      \\% (16)  &               --   & 
''        &  --   & --   & --    &  --    & --    &   7.2  &(15)   & 0.6   &(15)  & --                      \\% --    &              --   & 
''        &  --   & --   & --    &  --    & --    &   6.6  &(15)   & 0.572 &(18)  & --                      \\% --    &              --   & 
''        &  --   & --   & --    &  --    & --    &  11.5  &(23)   & --    & --   & --                      \\% --    &    --   & 
                                                                                                                                         
TWA-05B   &  M8.5 & (08) & 20.40 & 15.8   & --    &   5.1  &(15)   & 0.3   &(15)  & CD-33d7795B             \\% (08)  &     (14) & 
''        &  --   & --   & --    & 15.80  & --    &  20.   &(09)   & --    & --   & --                      \\% (16)  &               --   & 
                                                                                                                                         
TWA-06    &  K7   & (08) & 11.62 &  9.94  & --    &   4.65 &(08)   & 0.56  &(08)  & TYC7183-1477-1          \\% (10)  &     (08) & 
''        &  --   & --   & 12.00 &  9.77  & --    &   3.4  &(23)   & --    & --   & --                      \\% (16)  &                 --   & 
                                                                                                                                         
TWA-07    &  M1   & (08) & 11.65 &  9.21  & --    &   4.95 &(08)   & 0.44  &(08)  & TYC7190-2111-1          \\% (10)  &     (08) & 
''        &  --   & --   & 11.06 &  9.10  & --    &   5.8  &(23)   & --    & --   & --                      \\% (16)  &                  --   & 
                                                                                                                                         
TWA-08A   &  M2   & (08) & 12.23 &  9.82  & --    &   7.34 &(08)   & 0.53  &(08)  & RXJ1132.7-2651          \\% (10)  &     (08) & 
''        &  M2Ve & (10) & --    &  9.61  & --    &  25.   &(10)   & 0.520 &(10)  & --                      \\% (16)  &                 --   & 
''        &  --   & --   & --    &  --    & --    &   8.0  &(23)   & --    & --   & --                      \\% --    &    --   & 
                                                                                                                                         
TWA-08B   &  M5   & (08) & --    & 11.48  & --    &  16.5  &(08)   & 0.56  &(08)  & --                      \\% (12)  &      (08) & 
''        &  --   & --   & --    &  --    & --    &  13.3  &(23)   & --    & --   & --                      \\% --    &    --   & 
                                                                                                                                         
TWA-09A$^*$&  K5   & (08) & 11.26 &  9.64  & 19.87 &   2.35 &(08)   & 0.46  &(08)  & CD-36d7429A             \\% (10)  &     (08) & 
''        &  --   & --   & 11.32 &  9.60  & --    &   3.2  &(11)   & 0.47  &(18)  & --                      \\% (16)  &              --   & 
''        &  --   & --   & --    &  --    & --    &   2.1  &(23)   & --    & --   & --                      \\% --    &     --   & 
                                                                                                                                         
TWA-09B$^*$& M1   & (08) & 14.00 & 11.42  & 19.87 &   5.01 &(08)   & 0.48  &(08)  & --                      \\% (10)  &     (08) & 
''        &  --   & --   & 14.10 & 11.45  & --    &   5.1  &(11)   & --    & --   & --                      \\% (16)  &      --   & 
''        &  --   & --   & --    &  --    & --    &   4.3  &(23)   & --    & --   & --                      \\% --    &    --   & 
                                                                                                                                         
\hline
\end{tabular}
$\,$\\%                                                                                                                      
%%%%
%%%%(10)= Brandeker et al.       (2003),    
%%%%
$^*$ Member?\\
(01)= Rucinski \& Krautter   (1983),
(02)= de la Reza et al.      (1989),       
(03)= Jura et al.            (1993),        
(04)= Stauffer et al.        (1995),
(05)= Soderblom et al.       (1996),
(06)= Hoff et al.            (1998),
(07)= Sterzik  et al.        (1999),      
(08)= Webb et al.            (1999),               
(09)= Neuhauser et al.       (2000),      
(10)= Torres et al.          (2000),      
(11)= Muzerolle, priv. comm  (2000),   
(12)= Zuckerman et al.       (2001),       
(13)= Gizis                  (2002),          
(14)= Song et al.            (2002),        
(15)= Mohanty et al.         (2003),       
(16)= Reid et al.            (2003),           
(17)= Song et al.            (2003),        
(18)= Torres et al.          (2003),        
(19)= Gizis et al.           (2004),       
(20)= Weinberger et al.      (2004),      
(21)= Zuckerman \&  Song     (2004),      
(22)= Scholz \& Jayawardhana (2006),
(23)= Jayawardhana et al.    (2006),
(24)= Scholz et al.          (2006),
(24)= this paper
\end{table*}                                      
                                                                                                                                         
\setcounter{table}{0}
\begin{table*}[ht]
\caption[]{Data for members of the  TW Hydrae association}
\small
\begin{tabular}{lccccccclcl}
TWA       &Sp.Type&  ref.&  V    & I      &Parallax&W(Halpha)&ref  &  W(Li)&  ref &  Name  \\
\hline
                                                                                                                                         
TWA-10    &  M2.5 & (08) & 12.96 & 10.49  & --    &   8.39 &(08)   & 0.46  &(08)  & 1RXSJ123504.4-413629    \\% (10)  &     (08) & 
''        &  --   & --   & 12.70 & 10.50  & --    &   7.1  &(11)   & --    & --   & --                      \\% (16)  &                       --   & 
''        &  --   & --   & --    &  --    & --    &   4.5  &(11)   & --    & --   & --                      \\% --    &                       --   & 
''        &  --   & --   & --    &  --    & --    &  13.6  &(23)   & --    & --   & --                      \\% --    &    --   & 
                                                                                                                                         
TWA-11A   &  A0   & (08) &  5.78 &  5.81  & 14.91 &  --    & --    & --    & --   & HR4796A                 \\% (03)  &      (14) & 
                                                                                                                                         
TWA-11B   &  M2.5 & (08) & 13.3  & 10.78  & 14.91 &   3.5  &(04)   & 0.550 &(04)  & HR4796B                 \\% (03)  &     (14) & 
''        &  --   & --   & 12.80 & 10.60  & --    &   3.5  &(23)   & --    & --   & --                      \\% (16)  &           --   & 
                                                                                                                                         
TWA-12    &  M2   & (07) & 12.85 & 10.49  & --    &   4.5  &(07)   & 0.530 &(07)  & RXJ1121.1-3845          \\% (10)  &     (07) & 
''        &  --   & --   & 13.60 & 11.35  & --    &   5.3  &(11)   & --    & --   & --                      \\% (16)  &                  --   & 
''        &  --   & --   & --    &  --    & --    &   4.8  &(23)   & --    & --   & --                      \\% --    &    --   & 

TWA-13A   &  M2e  & (07) & 11.46 & --     & --    &   4.0  &(07)   & 0.650 &(07)  & RXJ1121.3-3447N         \\%  --   &     (07) & 
''        &  --   & --   & 12.10 & 10.10  & --    &   4.0  &(11)   & 0.650 &(18)  & --                      \\% (16)  &                   --   & 
''        &  --   & --   & --    &  --    & --    &   3.0  &(23)   & --    & --   & --                      \\% --    &    --   & 
                                                                                                                                         
TWA-13B   &  M1e  & (07) & 12.00 & --     & --    &   2.8  &(07)   & 0.550 &(07)  & RXJ1121.3-3447S         \\%  --   &     (07) & 
''        &  --   & --   & 12.40 & 10.10  & --    &   1.8  &(11)   & 0.570 &(18)  & --                      \\% (16)  &                   --   & 
''        &  --   & --   & --    &  --    & --    &   3.0  &(23)   & --    & --   & --                      \\% --    &    --   & 
                                                                                                                                         
TWA-14    &  M0   & (12) & 13.80 & 10.95  & --    &   8.2  &(12)   & 0.600 &(12)  & 1RXSJ111325.1-4523 44   \\% (16)  &     (12) & 
''        &  --   & --   & --    &  --    & --    &  12.5  &(11)   & --    & --   & --                      \\% --    &                        --   & 
''        &  --   & --   & --    &  --    & --    &  10.7  &(23)   & --    & --   & --                      \\% --    &    --   & 
                                                                                                                                         
TWA-15A   &  M1.5 & (12) & 14.10 & 11.94  & --    &   9.0  &(12)   & 0.650 &(12)  & 1RXSJ123420.1-4815 14   \\% (12)  &     (12) & 
''        &  --   & --   & --    & 11.94  & --    &   9.2  &(11)   & --    & --   & --                      \\% (16)  &                        --   & 
''        &  --   & --   & --    &  --    & --    &   8.8  &(23)   & --    & --   & --                      \\% --    &    --   & 
                                                                                                                                         
TWA-15B   &  M2   & (12) & 14.00 & 11.94  & --    &  10.5  &(12)   & 0.540 &(12)  & --                      \\% (12)  &     (12) & 
''        &  --   & --   & --    & 11.81  & --    &   8.6  &(23)   & --    & --   & --                      \\% (16)  &      --   & 
                                                                                                                                         
TWA-16    &  M1.5 & (12) & 12.30 & 10.17  & --    &   4.0  &(12)   & 0.360 &(12)  & 1RXSJ123456.1-453808    \\% (12)  &     (12) & 
''        &  --   & --   & --    & 10.17  & --    &   3.6  &(11)   & --    & --   & --                      \\% (16)  &                      --   & 
''        &  --   & --   & --    &  --    & --    &   4.0  &(23)   & --    & --   & --                      \\% --    &    --   & 
                                                                                                                                         
TWA-17    &  K5   & (12) & 12.70 & 10.78  & --    &   2.5  &(12)   & 0.490 &(12)  & 1RXSJ132046.5-461139    \\% (12)  &     (12) & 
''        &  --   & --   & --    & 10.78  & --    &   2.7  &(11)   & --    & --   & --                      \\% (16)  &                       --   & 
''        &  --   & --   & --    &  --    & --    &   3.2  &(23)   & --    & --   & --                      \\% --    &    --   & 
                                                                                                                                         
TWA-18    &  M0.5 & (12) & 12.90 & 10.92  & --    &   3.5  &(12)   & 0.420 &(12)  & 1RXSJ132137.0-442133    \\% (12)  &     (12) & 
''        &  --   & --   & --    & 10.92  & --    &   3.4  &(11)   & --    & --   & --                      \\% (16)  &                       --   & 
''        &  --   & --   & --    &  --    & --    &   3.3  &(23)   & --    & --   & --                      \\% --    &    --   & 
                                                                                                                                         
TWA-19A$^*$& G5   & (12) &  9.07 & --     &  9.62 &  -2.6  &(12)   & 0.190 &(12)  & HD102458                \\%  --   &     (12) & 
''        &  --   & --   &  9.14 & --     &  --   &  -0.57 &(14)   & 0.189 &(14)  & --                      \\%  --   &           --   & 
''        &  --   & --   &  9.10 &  8.40  &  --   &  --    & --    & --    & --   & --                      \\% (16)  &                    --   & 
                                                                                                                                         
TWA-19B$^*$& K7   & (12) & 11.90 & 10.21  &  9.62 &   2.8  &(12)   & 0.400 &(12)  & 1RXSJ114724.3-495250    \\% (16)  &    (12) & 
''        &  --   & --   & --    &  --    & --    &   2.2  &(23)   & --    & --   & --                      \\% --    &    --   & 
TWA-21    &  K3/4 & (21) &  9.8  & --     & --    &   0.0  &(17)   & 0.290 &(17)  & TYC8599-0697            \\%  --   &      (17) & 
                                                                                                                                         
TWA-22    &  M5   & (21) & 14.0  & --     & --    &   8.3  &(17)   & 0.510 &(17)  & SSS101726.7-535428      \\%  --   &     (17) & 
''        &  --   & --   & --    &  --    & --    &  11.5  &(23)   & --    & --   & --                      \\% --    &    --   & 
                                                                                                                                         
TWA-23    &  M1   & (20) & 12.67 & --     & --    &   2.2  &(17)   & 0.460 &(17)  & SSS120727.4-324700      \\%  --   &     (17) & 
''        &  --   & --   & --    &  --    & --    &   2.4  &(23)   & --    & --   & --                      \\% --    &    --   & 
                                                                                                                                         
TWA-24    &  K3   & (21) & 12.26 & --     & --    &   0.5  &(17)   & 0.337 &(17)  & TYC8644-0802            \\%  --   &      (17) & 
''        &  --   & --   & --    &  --    & --    &   0.3  &(23)   & --    & --   & --                      \\% --    &    --   & 
                                                                                                                                         
TWA-25    &  M0   & (20) & 11.36 & --     & --    &   1.8  &(17)   & 0.494 &(17)  & TYC7760-0283            \\%  --   &      (17) & 
''        &  --   & --   & --    &  --    & --    &   2.4  &(23)   & --    & --   & --                      \\% --    &    --   & 
                                                                                                                                         
TWA-26    &  M8   & (13) & 20.50 & 15.88  & --    & 300.   &(13)   & 0.5   &(15)  & 2MJ1207334-393254       \\% (19)  &     (13) & 
''        &  --   & --   & --    & 16.00  & --    &  42.   &(19)   & 0.4   &(24)  & --                      \\% (16)  &             --   & 
''        &  --   & --   & --    &  --    & --    &  27.7  &(15)   & 0.51  &(25)  & --                      \\% --    &             --   & 
''        &  --   & --   & --    &  --    & --    &12--387 &(22)   & --    & --   & --                      \\% --    &    --   & 
''        &  --   & --   & --    &  --    & --    &  44.7  &(25)   & --    & --   & --                      \\% --    &    --   & 
                                                                                                                                         
TWA-27    &  M8   & (13) & 20.20 & 15.70  & --    &  10.   &(13)   & 0.5   &(15)  & 2MJ1139511-315921       \\% (16)  &     (13) & 
''        &  --   & --   & --    &  --    & --    &   7.3  &(15)   & 0.62  &(25)  & --                      \\% --    &          --   & 
''        &  --   & --   & --    &  --    & --    &  10.2  &(25)   & --    & --   & --                      \\% --    &    --   & 
\hline
\end{tabular}
$\,$\\%                                                                                                                      
%%%%
%%%%(10)= Brandeker et al.       (2003),    
%%%%
$^*$ Member?\\
(01)= Rucinski \& Krautter   (1983),
(02)= de la Reza et al.      (1989),       
(03)= Jura et al.            (1993),        
(04)= Stauffer et al.        (1995),
(05)= Soderblom et al.       (1996),
(06)= Hoff et al.            (1998),
(07)= Sterzik  et al.        (1999),      
(08)= Webb et al.            (1999),               
(09)= Neuhauser et al.       (2000),      
(10)= Torres et al.          (2000),      
(11)= Muzerolle, priv. comm  (2000),   
(12)= Zuckerman et al.       (2001),       
(13)= Gizis                  (2002),          
(14)= Song et al.            (2002),        
(15)= Mohanty et al.         (2003),       
(16)= Reid et al.            (2003),           
(17)= Song et al.            (2003),        
(18)= Torres et al.          (2003),        
(19)= Gizis et al.           (2004),       
(20)= Weinberger et al.      (2004),      
(21)= Zuckerman \&  Song     (2004),      
(22)= Scholz \& Jayawardhana (2006),
(23)= Jayawardhana et al.    (2006),
(24)= Scholz et al.          (2006),
(24)= this paper

\end{table*}

\end{document}